\documentclass[onecolumn,amsmath,showpacs,nofootinbib,12pt]{revtex4-2}
\usepackage{graphicx}
\usepackage{dcolumn}
\usepackage{bm}
\usepackage{color} 
\usepackage{slashed}
\begin{document}
\newcommand{\hs}{\hspace*{0.5cm}}
\newcommand{\vs}{\vspace*{0.5cm}}
\newcommand{\be}{\begin{equation}}
\newcommand{\ee}{\end{equation}}
\newcommand{\bea}{\begin{eqnarray}}
\newcommand{\eea}{\end{eqnarray}}
\newcommand{\ben}{\begin{enumerate}}
\newcommand{\een}{\end{enumerate}}
\newcommand{\bde}{\begin{widetext}}
\newcommand{\ede}{\end{widetext}}
\newcommand{\nn}{\nonumber}
\newcommand{\crn}{\nonumber \\}
\newcommand{\Tr}{\mathrm{Tr}}
\newcommand{\non}{\nonumber}
\newcommand{\noi}{\noindent}
\newcommand{\al}{\alpha}
\newcommand{\la}{\lambda}
\newcommand{\bet}{\beta}
\newcommand{\ga}{\gamma}
\newcommand{\va}{\varphi}
\newcommand{\om}{\omega}
\newcommand{\pa}{\partial}
\newcommand{\+}{\dagger}
\newcommand{\fr}{\frac}
\newcommand{\bc}{\begin{center}}
\newcommand{\ec}{\end{center}}
\newcommand{\Ga}{\Gamma}
\newcommand{\de}{\delta}
\newcommand{\De}{\Delta}
\newcommand{\ep}{\epsilon}
\newcommand{\varep}{\varepsilon}
\newcommand{\ka}{\kappa}
\newcommand{\La}{\Lambda}
\newcommand{\si}{\sigma}
\newcommand{\Si}{\Sigma}
\newcommand{\ta}{\tau}
\newcommand{\up}{\upsilon}
\newcommand{\Up}{\Upsilon}
\newcommand{\ze}{\zeta}
\newcommand{\ps}{\psi}
\newcommand{\Ps}{\Psi}
\newcommand{\ph}{\phi}
\newcommand{\vph}{\varphi}
\newcommand{\Ph}{\Phi}
\newcommand{\Om}{\Omega}
\newcommand{\AdrHEPC}{Phenikaa Institute for Advanced Study, Phenikaa University, Nguyen Trac, Duong Noi, Hanoi 100000, Vietnam}

\title{Implications of a dark grand unification} 
\author{Phung Van Dong} 
\email{dong.phungvan@phenikaa-uni.edu.vn (corresponding author)}
\affiliation{\AdrHEPC} 
\author{Do Thi Huong}
\email{dthuong@iop.vast.vn}
\affiliation{Institute of Physics, Vietnam Academy of Science and Technology, 10 Dao Tan, Giang Vo, Hanoi 100000, Vietnam}

\date{\today}

\begin{abstract}

Given that dark matter and normal matter are nontrivially unified in a grand unified theory, called dark grand unification, we derive novel residual theories at low energy explaining dark matter and neutrino mass. The first chain of which is $E_6\to SU(3)_C\otimes SU(3)_L\otimes SU(3)_R$, which contains a matter parity by itself stabilizing a dark matter candidate and producing neutrino mass via a seesaw. The second chain is $\mathrm{Trinification}\to SU(3)_C \otimes SU(3)_L\otimes U(1)_X\otimes U(1)_N$, which results in a novel family-universal 3-3-1-1 model, opposite to the normal 3-3-1-1 (or corresponding 3-3-1) model. Surprisingly, it is a variant of both minimal 3-3-1 model and 3-3-1 model with right-handed neutrinos since both $e_R$ and $\nu_R$ are located at the bottoms of lepton triplets. Since this universal 3-3-1-1 model is properly embedded in the trinification, the above matter parity works governing dark matter stability as well as suppressing unwanted fermion mixings. Further, neutrino masses are naturally generated by a canonical seesaw combined with a scotogenic scheme.            
                  
\end{abstract} 

\maketitle

\section{Introduction}

The existence of dark matter, which makes up most of the mass of galaxies and galaxy clusters, is one of the biggest challenges in the modern physics \cite{bertone,arcadi}. The standard model, which is our current understanding of the matter content of the universe, cannot solve this issue~\cite{pdg}. A number of the well-motivated extensions of the standard model have been proposed so far, providing a dark matter candidate, for instance, supersymmetry \cite{susy1,susy2,susy3,susy4}, extradimensions \cite{extra1,extra2,extra3,extra4}, little Higgs model \cite{little1,little2,little3,little4}, minimal dark matter \cite{minidm}, scotogenic setup \cite{scoto1,scoto2,scoto3}, and dark photon model \cite{dp1,dp2,dp3}. Obviously, many of such candidates according to the extensions are well-compelling. However, this work would not go into details of any of them because there have not experimentally indicated about properties of a candidate. Alternatively, we suggest that at high energy the fundamental interactions and particles are unified, which would include the dark interaction and matter too. Conversely, a dark grand unification is possible for which its breaking properly leads to a correct dark interaction and matter at low energy, which is not arbitrarily imposed as in the old approaches.   

Before searching for a dark grand unification, it is noted that several non-Abelian gauge symmetries \cite{d1,d1e1,d1e2,d1e3,d1e5,d1e6,e7,e8,e9,d2,d3} may provide a dark matter candidate manifestly from the gauge principle, similar to the case of dark photon theory with Abelian gauge symmetry. The traditional grand unified theories may unify all particle forces such as $SU(5)$ \cite{su5}, $SO(10)$ \cite{so10,so10p}, and even Pati-Salam model \cite{ps} but they cannot contain any dark matter candidate, somewhat similar to the case of the standard model. This is because the extra force or particle is indeed a leptoquark, which is not color neutral as expected for dark matter. Further, all of them directly couple to normal matter, which would decay. However, a class of grand unification based upon exceptional groups $E_{6}$ \cite{e68} and trinification \cite{trinifi1} may be a suitable choice, since their matter representations contain rich new fields responsible for dark matter. In this work, we examine the trinification as the minimal framework among them.

The trinification has been extensively studied in \cite{trinifi2,trinifi3,trinifi4,trinifi5,trinifi6,trinifi7,trinifi8}. Possibility of dark matter existence in this kind of model has been discussed in \cite{trinifi9p}. However, one need give up a unification as the symmetry is flipped. The option of dark matter within the minimal trinification framework has been reinterpreted in \cite{trinifi9}. There, a trinification parity which depends on the generalized isospins has been examined, making the vector dark matter stable. The trinification parity is identical to the matter parity in \cite{trinifi9p} if the flipped is suppressed. In this work, we reinvestigate the trinification model and show that the trinification parity is indeed a matter parity, analogous to that in supersymmetry. Hence, the trinification at low energy properly addresses dark matter and neutrino mass \cite{kajita,mcdonald}. It is both a residual theory of and implied by the dark grand unification $E_6$ at very high scale. 

In alternative perspective, we indicate that the trinification at high energy acts as a dark grand unification. It is reduced to (i.e., embedded in by) a novel 3-3-1-1 model before it is broken to the standard model. This 3-3-1-1 model has not been recognized before, providing neutrino masses and dark matter candidates manifestly. Looking into the literature, the corresponding 3-3-1 version and a variant of it were proposed in \cite{331u1,331u2,331u3}. However, properly embedding in the trinification requires a (larger) intermediate 3-3-1-1 symmetry instead at the same energy regime for which the last $U(1)$ factor as enclosed especially determines $B-L$ and matter parity responsible for dark matter stability and neutrino mass generation. Additionally, such 3-3-1-1 and 3-3-1 models are family universal, hence the FCNCs associated with $Z'$ bosons are suppressed. As a matter of fact, it is well established that the normal 3-3-1 model \cite{331m1,331m2,331m3,331r0,331r1,331r2,331r3} is not family universal, opposite to the standard model, due to anomaly cancellation. This gives rise to dangerous FCNCs mediated by the $Z'$ boson despite the fact that the family number is fixed as three. Additionally, it does not simply embed the normal 3-3-1 model in a grand unification, unlike the standard model and other extensions. The universal 3-3-1 (exactly 3-3-1-1) model as proposed appropriately resolves all the issues, i.e. implied by grand unification, dark matter stability, and small neutrino mass. 

The rest of this work is organized as follows. In Sec. \ref{model}, we prove that as reduced from $E_6$ grand unification, the trinification possesses a matter parity in itself, which classifies types of matter within gauge multiplets, making the lightest odd particle stable. In Sec. \ref{u331}, we propose the universal 3-3-1-1 model, which is reduced from the trinification and containing a reasonable minimal amount of dark matter candidates. In Sec. \ref{pheno}, we investigate the phenomenology of the universal 3-3-1-1 model, such as fermion, gauge boson, and scalar mass spectra, dark matter observables, and collider signals. Lastly, we summarize our results and conclude this work in Sec. \ref{con}.      
    
\section{\label{model} Trinification as reduced from $E_6$}

The history of grand unified theories is intriguing. The idea by which the standard model gauge symmetry is a maximal subgroup of $SU(5)$, which combined with a $U(1)$ is a maximal subgroup of $SO(10)$, translates in a compelling and systematic way until the largest exception group. The elegant breaking chain $E_8\supset E_7\supset E_6 \supset SO(10)\supset SU(5)\supset \mathrm{SM}$ together with $E_8$ played in string theory perhaps reveal a deep link between particle physics and quantum gravity. Nevertheless, because the representations of $E_{7,8}$ are real, they are not useful for unification. Additionally, $SU(5)$ and $SO(10)$ do not manifestly contain dark matter. $E_6$ does turn out to be interesting, a candidate for dark grand unification. $E_6$ has $SU(3)\otimes SU(3) \otimes SU(3)$ as a maximal subgroup, where the nature of each factor group is detailed below. The fundamental and adjoint representations of $E_6$ are 27- and 78-dimensional, respectively. They decompose into $SU(3)\otimes SU(3) \otimes SU(3)$ such as 
\bea 27 &=& (3,3,1)\oplus (3^*,1,3^*)\oplus (1,3^*,3),\\
78 &=& (8,1,1)\oplus (1,8,1)\oplus (1,1,8)\oplus (3^*,3,3) \oplus (3, 3^*,3^*).\eea A real scalar (rank 2 tensor) representation with appropriate large VEV will break $E_6$ to $SU(3)\otimes SU(3) \otimes SU(3)$, making leptoquark bosons $(3^*,3,3)$ and $(3, 3^*,3^*)$ heavy, as integrated out \cite{babu}. What remains at low energy is a trinification, specified below. 

That said, the trinification gauge symmetry is given by 
\be SU(3)_C\otimes SU(3)_L\otimes SU(3)_R,\ee where the first factor denotes the color group, as usual, while the second and last factors both contain the electroweak symmetry and transform nontrivially left-handed and right-handed fermions, respectively. The fermion content, which is anomaly free, is given by 
\bea Q_{L} &=& \begin{pmatrix} u\\ d\\ D \end{pmatrix}_L\sim (3,3,1),\\
Q_{R} &=& \begin{pmatrix} u\\ d\\ D \end{pmatrix}_R\sim (3,1,3),\\
\Psi_{L} &=& \begin{pmatrix} E^0 & E^- & e\\ 
E^+ & E^{c0} & -\nu\\ 
e^c& -\nu^c& N \end{pmatrix}_L\sim (1,3^*,3).\label{eq11}\eea Here, the values in parentheses, say $(C,L,R)$, denote the quantum numbers according to the groups $(SU(3)_C, SU(3)_L, SU(3)_R)$, respectively. A triplet (antitriplet) of $SU(3)_{L,R}$ is decomposed into $3=2\oplus 1$ ($3^*=2^*\oplus 1$) under the subgroup $SU(2)_{L,R}$. Hence, it is clear that $(u,d)^T$ transforms as 2, while $(e,-\nu)^T$ transform as $2^*$, analogous to the standard model. We denote $f^c_L\equiv (f^c)_L=(f_R)^c=C\bar{f}_R^T$ to be the charge conjugation of the relevant right-handed fermion. It is noted that $\nu_L,e,u,d$ are usual fermions, while $\nu^c_L,N,E,D$ are new fermions. Above, the fermion representations correspond to a family, for which the family index is suppressed. Similarly, the color index is suppressed too.

The conservation and additive nature of electric charge demand that the electric charge is combined of neutral generators of $SU(3)_L\otimes SU(3)_R$, such as $Q=a T_{3L} + b T_{3R}+cT_{8L}+dT_{8R}$. Acting this operator on the fermion representations, with known electric charge of usual particles, we obtain $a=b=1$ and $c=d=1/\sqrt{3}$, thus 
\be Q= T_{3L} + T_{3R}+\fr{1}{\sqrt{3}}(T_{8L}+ T_{8R}).\ee The left-right symmetry originating from the trinification is still reflected at this step and matches the baryon-minus-lepton number, say  \be B-L=\fr{2}{\sqrt{3}}(T_{8L}+T_{8R}),\ee as in the usual left-right symmetric model. Further, comparing to that in the standard model, i.e. $Q=T_{3L}+Y$, the hypercharge is given by  
\be Y=T_{3R}+\fr 1 2 (B-L).\ee

To break the trinification symmetry and produce fermion masses properly, we introduce \ben 
\item Three scalar bi-triplets $(\phi_i)_a^x\sim (1,3,3^*)$, where $i=1,2,3$ labels a bi-triplet, while $a$ and $x$ denote $SU(3)_L$ and $SU(3)_R$ indices, respectively. Such a bi-triplet (i.e., the subscript $i$ is suppressed) takes the form,
\be \phi_a^x =\begin{pmatrix} 
\phi^0_{11}& \phi^+_{12} & \phi^+_{13}\\
\phi^-_{21} & \phi^0_{22} & \phi^0_{23}\\
\phi^-_{31} & \phi^0_{32} & \phi^0_{33}\end{pmatrix} \sim (1,3,3^*),\label{sbt}\ee which couples to both quarks as $\bar{Q}^a_L \phi_a^x Q_{Rx}$ and leptons as $\ep^{xyz}\ep_{abc}\Psi^a_{Lx}\Psi^b_{L y} (\phi^*)^c_z$, where $b,c$ like $a$ and $y,z$ like $x$ are $SU(3)_L$ and $SU(3)_R$ indices, respectively.\item One scalar bi-sextet $\chi_{ab}^{xy}\sim (1,6,6^*)$, where $a,b$ and $x,y$ correspondingly stand for $SU(3)_L$ and $SU(3)_R$ indices, as mentioned. The bi-sextet couples only to leptons (not quarks) such as $\Psi^a_{Lx}\Psi^b_{Ly} \chi^{xy}_{ab}$, i.e. the two elements/leptons of (\ref{eq11}) at rows $a,b$ and columns $x,y$ are coupled by $\chi^{xy}_{ab}$, respectively\een    

Notice that $SU(3)_L\supset SU(2)_L \otimes U(1)_{T_{8L}}$ and $SU(3)_R\supset SU(2)_R \otimes U(1)_{T_{8R}}$. And, we assume that the trinification is broken directly to the standard model (i.e., no intermediate physical phase is presented). The first stage of the trinification breaking down to the standard model is determined by\ben 
\item $SU(3)_L\otimes SU(3)_R \to SU(2)_L\otimes SU(2)_R\otimes U(1)_{B-L}$ is induced by a new physics scale ($\sim w$), where $B-L=(2/\sqrt{3})(T_{8L}+T_{8R})$ as given conserves $w$. 
\item Simultaneously, $SU(2)_R\otimes U(1)_{B-L}\to U(1)_Y\otimes M_P$ is broken by the right-handed neutrino mass scale ($\sim \La$), where $Y=T_{3R}+\fr 1 2 (B-L)$ is as usual, while $M_P=(-1)^{3(B-L)}$ is the matter parity, conserved after breaking.\een It is used to form $M_P=(-1)^{3(B-L)+2s}$, which is multiplied by the spin parity $(-1)^{2s}$ conserved by the Lorentz symmetry (the derivation of the matter parity can be done similarly to the next section, without loss of generality). Including the second stage of the electroweak breaking by $u,v$ scales, the scheme of the trinification breaking is given by 
\bc \begin{tabular}{c}
$SU(3)_C\otimes SU(3)_L\otimes SU(3)_R$ \\
$\downarrow w,\La $\\
$SU(3)_C\otimes SU(2)_L\otimes U(1)_Y\otimes M_P$\\
$\downarrow u,v$\\
$SU(3)_C\otimes U(1)_Q\otimes M_P$ 
\end{tabular}\ec       

\begin{table}[h]
\bc
\begin{tabular}{r|cccccccccccc}
\hline\hline
Particle & $\nu$ & $e$ & $u$ & $d$ & $N$ & $E$ & $D$ & $G$ & $A_{3,8}$& $W$ & $X$ & $Y$\\
\hline 
$B-L$ & $-1$ & $-1$ & $\fr 1 3 $ & $\fr 1 3 $ & 0 & 0 & $-\fr{2}{3}$ & 0 & 0 & 0 & 1 & 1\\
$M_P$ & $+$ & $+$ & $+$ & $+$ & $-$ & $-$ & $-$ & $+$ & $+$ & $+$ & $-$ & $-$\\
\hline \hline
\end{tabular}
\caption[]{\label{tab1} $B-L$ number and matter parity of fermions and gauge bosons in the trinification.}
\ec
\end{table} 
\begin{table}[h]
\bc
\begin{tabular}{r|cccccccccccccccc}
\hline\hline
Particle & $\phi^0_{11}$ & $\phi^+_{12}$ & $\phi^-_{21}$ & $\phi^0_{22}$ & $\phi^0_{33}$ & $\phi^+_{13}$ & $\phi^0_{23}$ & $\phi^-_{31}$ & $\phi^0_{32}$ & $\chi_{\al \beta}^{\ga \de}$ & $\chi^{33}_{33}$ & $\chi^{\ga 3/3\de}_{\al 3/3\beta}$ & $\chi^{33}_{\al \beta}$ & $\chi^{\ga \de}_{33}$ & $\chi^{\ga 3/3\de}_{\al \beta/33}$ & $\chi^{\ga\de/33}_{\al 3/3\beta}$\\
\hline 
$B-L$ & $0$ & $0$ & 0 & 0 & 0 & $1 $ & $1 $ & $-1$ & $-1$ & 0 & 0 & 0 & $2$ & $-2$ & 1 & $-1$\\
$M_P$ & $+$ & $+$ & $+$ & $+$ & $+$ & $-$ & $-$ & $-$ & $-$ & $+$ & $+$ & $+$ & $+$ & $+$ & $-$ & $-$ \\
\hline \hline
\end{tabular}
\caption[]{\label{tab2} $B-L$ number and matter parity of scalar particles in the trinification.}
\ec
\end{table}
The $B-L$ charge and $M_P$ value of the model particles are collected in Tables \ref{tab1} and \ref{tab2}, where we assign $\al,\beta=1,2$ and $\ga,\de=1,2$ to be the first two values of $a,b$ and $x,y$, which belong to the $SU(2)_{L,R}$ subgroups, respectively. Additionally, apart from the gluon field $G$ coupled to $SU(3)_C$, as usual, the adjoint gauge fields associated with the gauge groups $SU(3)_{L,R}$ are given, respectively, by \be A_{L,R} = A_{nL,R} T_{nL,R}=\begin{pmatrix}\fr 1 2 A_3+\fr{1}{2\sqrt{3}} A_8 & \fr{1}{\sqrt{2}}W^+ & \fr{1}{\sqrt{2}}X^+\\
\fr{1}{\sqrt{2}}W^- & -\fr 1 2 A_3+\fr{1}{2\sqrt{3}} A_8  & \fr{1}{\sqrt{2}}Y^0\\
\fr{1}{\sqrt{2}}X^- & \fr{1}{\sqrt{2}}Y^{0*}& -\fr{1}{\sqrt{3}} A_8  \end{pmatrix}_{L,R},\ee where $W^\pm=(A_1\mp i A_2)/\sqrt{2}$, $X^\pm=(A_4\mp i A_5)/\sqrt{2}$, and $Y^{0,0*}=(A_6\mp i A_7)/\sqrt{2}$. In other words, the particle multiplets have matter parity, 
\bea &&Q_{L} = \begin{pmatrix} +\\ +\\ - \end{pmatrix}_L,\hs Q_{R} = \begin{pmatrix} +\\ +\\ - \end{pmatrix}_R,\hs \Psi_{L} = \begin{pmatrix} - & - & +\\ 
- & - & +\\ 
+ & +& - \end{pmatrix}_L,\\ 
&& A_{L,R}= \begin{pmatrix} + & + & -\\ 
+ & + & -\\ 
- & -& + \end{pmatrix}_{L,R},\hs \phi = \begin{pmatrix} + & + & -\\ 
+ & + & -\\ 
- & -& + \end{pmatrix},\hs \chi = \begin{pmatrix} + & + & -\\ 
+ & + & -\\ 
- & -& + \end{pmatrix}, \eea where the tensor $\chi$ is displayed in 2D as $\chi^X_A$, which summarizes over subindex pairs, namely $A=\{\al\beta,33,\al 3/3\beta\}$ and $X=\{\ga\de,33,\ga 3/3\de\}$. This proves that normal fields and dark fields are unified in gauge-symmetry multiplets of the trinification. 

Remarks are given in order
\ben 
\item The matter parity conservation demands that $\phi^0_{23}$, $\phi^0_{32}$, $\chi^{\ga 3/3\de}_{\al\beta/33}$, and $\chi^{\ga\de/33}_{\al3/3\beta}$, which connect a usual fermion to a new fermion, have vanished vacuum expectation values (VEVs), although a number of them are electrically neutral. Additionally, the electric charge conservation demands that the electrically-charged scalars cannot have nonzero VEVs despite that their matter parity may be even. In other words, the fields $\phi_x$ have only a VEV of type $\langle \phi_x\rangle =\fr{1}{\sqrt{2}}\mathrm{diag}(u,v,w)_x$ for $x=1,2,3$. We can work in a basis so that the vacuum alignments are canonically orthogonalized, i.e. 
\be \langle \phi_1\rangle =\mathrm{diag}(u/\sqrt{2},0,0),\hs \langle \phi_2\rangle =\mathrm{diag}(0,v/\sqrt{2},0),\hs \langle \phi_3\rangle =\mathrm{diag}(0,0,w/\sqrt{2}).\ee Further, $\chi$ develops a VEV whose nonzero elements are 
\be \langle \chi^{33}_{33} \rangle=\La_1/\sqrt{2},\hs \langle \chi^{22}_{33} \rangle=\La_2/\sqrt{2},\hs\langle \chi^{22}_{22} \rangle=\La_3/\sqrt{2},\ee such that the left-right symmetry is broken by $\La=\{\La_1,\La_2,\La_3\}$, besides the conditions of matter parity and electric charge conservations, as mentioned. Here, like $u,v,w$, $\La_{1,3}$ conserves $B-L$. By contrast, $\La_2$ breaks $B-L$ by two units, defining the matter parity $M_P$. Let us impose $\La\sim w\gg u,v$ for consistency with the standard model.   
\item The matter parity conservation suppresses a potential mixing between $d$ and $D$ quarks, which have opposite matter parity values, which otherwise leads to dangerous FCNCs. 
\item Last, but not least, the lightest matter-parity odd particle, which may be a gauge field $Y^0$, one of fermions $N$ and $E^0$, or one of scalars $\phi^0_{23}$, $\phi^0_{32}$, $\chi^{\ga 3/3\de}_{\al\beta/33}$, and $\chi^{\ga\de/33}_{\al3/3\beta}$, is stabilized by the matter parity responsible for dark matter. 
\een

It is shown that neutrino mass is generated via a canonical seesaw, $m_\nu\sim u^2/\La$, to be naturally small. New leptons $(E,N)$ and quarks ($D$) gain large masses at $w,\La$ scale, similar to the Majorana mass of $\nu_R$. Particularly, considering the gauge boson candidate ($A_{7R}$) as dark matter, the dark matter experiment and collider indicate the new physics scale at TeV \cite{trinifi9}. Obviously, the trinification at low energy is a framework for dark matter and neutrino mass by itself and is manifestly unified at GUT scale, named $E_6$ dark grand unification. 

Although the particle scenarios of dark matter, say fermion type (3 singlets $N$'s, 3 doublets $E^0$'s), scalar type (3 singlets $\phi^0_{32}$'s, 3 doublets $\phi^0_{23}$'s and a lot of doublets in bi-sextets), and vector type (2 doublets $Y^0_{L,R}$), are worth exploring, we would not examine this model further. Instead, we will investigate the trinification breaking at very high scale, simultaneously looking for an intermediate new physics phase---the universal 3-3-1-1 model, whose number of vector and scalar candidates are significantly reduced to a predictive level.           

\section{\label{u331} Universal 3-3-1-1 model as reduced from trinification}

As mentioned, the trinification gauge symmetry is given by
\be SU(3)_C\otimes SU(3)_L\otimes SU(3)_R,\ee in which the electric charge and baryon-minus-lepton number are embedded in the gauge symmetry, such as \bea && Q= T_{3L} + T_{3R}+\fr{1}{\sqrt{3}}(T_{8L}+ T_{8R}),\\ && B-L=\fr{2}{\sqrt{3}}(T_{8L}+T_{8R}),\eea respectively. Now that, in contrast to the previous model, we assume the trinification to be an independent dark grand unification, broken at a very high scale.\footnote{Notice that this grand unification scale may be radically low, but still much beyond TeV, since the proton decay does not occur in the trinification.} There are two ways of symmetry breaking of interest, as shown below.
 
Because of $SU(3)_L\supset SU(2)_L \otimes U(1)_{T_{8L}}$ and $SU(3)_R\supset SU(2)_R \otimes U(1)_{T_{8R}}$, the trinification gauge symmetry is broken down to the standard model through intermediate groups,
$SU(3)_C\otimes SU(3)_L\otimes SU(3)_R\to SU(3)_C\otimes SU(2)_L\otimes SU(2)_R \otimes U(1)_{T_{8L}}\otimes U(1)_{T_{8R}}\to SU(3)_C\otimes SU(2)_L\otimes SU(2)_R \otimes U(1)_{B-L}\to SU(3)_C\otimes SU(2)_L \otimes U(1)_Y$, where $Y=T_{3R}+\fr 1 2 (B-L)$. This type of breaking requires a bi-triplet $\phi_3$ that has a VEV $w$ much beyond TeV scale (or $SU(2)_R\otimes U(1)_{B-L}$ scale). All the dark fields that are odd under matter parity are heavy at $w$ scale, as integrated out. In other words, this case does not provide any dark matter candidate as contributing negligibly to dark matter observables. Hence, the intermediate new physics phase is just the left-right symmetric model, as often studied and skipped. 

Alternatively, because of $SU(3)_R\supset U(1)_{T_{3R}} \otimes U(1)_{T_{8R}}$, the trinification gauge symmetry is broken to the standard model via intermediate groups, $SU(3)_C\otimes SU(3)_L\otimes SU(3)_R\to
SU(3)_C\otimes SU(3)_L\otimes U(1)_{T_{3R}}\otimes U(1)_{T_{8R}}\to SU(3)_C\otimes SU(3)_L\otimes U(1)_X \otimes U(1)_N\to SU(3)_C\otimes SU(2)_L \otimes U(1)_Y$. Here, the 3-3-1-1 symmetry, i.e.
\be SU(3)_C\otimes SU(3)_L\otimes U(1)_X \otimes U(1)_N, \ee arising from the trinification by conveniently rearranging the $U(1)$'s charges, such as 
\be X=T_{3R}+\fr{1}{\sqrt{3}}T_{8R},\ee which determines the electric charge $Q=T_{3L}+\fr{1}{\sqrt{3}}T_{8L}+X$, thus the hypercharge $Y=\fr{1}{\sqrt{3}}T_{8L}+X$, identical to that in the 3-3-1 model, and \be N=\fr{2}{\sqrt{3}}T_{8R},\ee which determines the baryon-minus-lepton number $B-L=\fr{2}{\sqrt{3}} T_{8L}+N$, analogous to that in the 3-3-1-1 model. This type of breaking acquires a scalar octet, \be S_{R}=S_{nR}T_{nR} =\begin{pmatrix} \fr{1}{2}S_3 +\fr{1}{2\sqrt{3}}S_8 & \fr{1}{\sqrt{2}}S^+_{12} & \fr{1}{\sqrt{2}} S^+_{13}\\
\fr{1}{\sqrt{2}} S^-_{12} & -\fr{1}{2}S_3 +\fr{1}{2\sqrt{3}}S_8 & \fr{1}{\sqrt{2}} S^0_{23}\\
\fr{1}{\sqrt{2}}S^-_{13} & \fr{1}{\sqrt{2}}S^{0*}_{23} & -\fr{1}{\sqrt{3}}S_8\end{pmatrix}_{R} \sim(1,1,8).\ee $S_R$ possessing a VEV $\langle S_R\rangle =\fr{1}{2\sqrt{2}} \mathrm{diag}(\Delta,-\Delta,0)$ when $S_{3R}$ develops a VEV $\langle S_{3R}\rangle = \Delta/\sqrt{2}$ breaks $SU(3)_R\to U(1)_{T_{3R}}\otimes U(1)_{T_{8R}}$.\footnote{See also \cite{dongloi} for another breaking by scalar octet.} This scheme of breaking requires $\Delta$ much beyond TeV scale (or $SU(3)_L\otimes U(1)_X\otimes U(1)_N$ scale, known as the 3-3-1-1 scale). A left-right symmetry may require another octet $S_L\sim (1,8,1)$ but have $\langle S_L\rangle =0$, while $\langle S_R\rangle$ is retained. It is noted that $S_R$ and $S_L$, even if a scalar bi-octet is imposed instead, do not couple to fermions. $S_R$ and $S_L$ like the non-Hermitian gauge bosons of $SU(3)_R$ are all heavy at $\Delta$ scale, hence being integrated out at the 3-3-1-1 scale. 

Under the decomposition $SU(3)_C\otimes SU(3)_L\otimes SU(3)_R\to SU(3)_C\otimes SU(3)_L\otimes U(1)_X \otimes U(1)_N$, the fermion representations are separated as  
\be Q_{L} = \begin{pmatrix} u\\ d\\ D \end{pmatrix}_L\sim (3,3,0,0),\ee
\bea && u_R \sim (3,1,2/3,1/3),\\ 
&& d_R\sim (3,1,-1/3,1/3),\\ 
&& D_{R}\sim (3,1,-1/3,-2/3),\eea
\be \psi_{L} = \begin{pmatrix}  e\\ 
 -\nu\\ 
 N \end{pmatrix}_L\sim (1,3^*,-1/3,-2/3),\ee 
\be \psi_{eL} = \begin{pmatrix} E^0 \\ 
E^+ \\ 
e^c \end{pmatrix}_L\sim (1,3^*,2/3,1/3),\ee 
\be \psi_{\nu L} = \begin{pmatrix}  E^- \\ 
 E^{c0} \\ 
-\nu^c \end{pmatrix}_L\sim (1,3^*,-1/3, 1/3),\ee which is anomaly free according to the 3-3-1-1 symmetry. Here, the values in parentheses, say $(C,L,X,N)$, denote the quantum numbers according to the 3-3-1-1 symmetry, in which $C$ and $L$ are given by representation dimensions, while $X$ and $N$ are given by such $U(1)$'s charges. The above fermion representations correspond to a family, while the other families are replicated. Each family has a new vectorlike quark $D$, a new vectorlike lepton doublet $(E^0,E^-)$, two new chiral leptons $\nu_R$ and $N_L$ under the standard model symmetry. It is noted that $N_L$ may have a right-handed partner $N_R$, which is a singlet under every symmetry as given. Strictly speaking, the presence of the new vectorlike lepton doublets makes the model family universal. They couple to $e_R$ and $\nu_R$ in a minimal way, similar to the minimal 3-3-1 model \cite{331m1,331m2,331m3} and the 3-3-1 model with right-handed neutrinos \cite{331r0,331r1,331r2,331r3}.      

According to the decomposition $SU(3)_C\otimes SU(3)_L\otimes SU(3)_R\to SU(3)_C\otimes SU(3)_L\otimes U(1)_X \otimes U(1)_N$, a scalar bi-triplet is separated into 
\bea &&\eta = \begin{pmatrix} \eta^0_1\\
\eta^-_2\\
\eta^-_3
\end{pmatrix}\sim (1,3,-2/3,-1/3),\\
&& \rho = \begin{pmatrix} \rho^+_1\\
\rho^0_2\\
\rho^0_3
\end{pmatrix}\sim (1,3,1/3,-1/3),\\ 
&&\varphi = \begin{pmatrix} \varphi^+_1\\
\varphi^0_2\\
\varphi^0_3
\end{pmatrix}\sim (1,3,1/3,2/3),\eea where the fields are renamed, as compared with (\ref{sbt}), for brevity.\footnote{Assuming only a scalar bi-triplet originating from the dark grand unification alive as separated at low energy, opposite to the low energy trinification version.} Additionally, the scalar bi-sextet is separated into six sextets under the decomposition. Opposite to the scalar bi-triplet whose separations all break the 3-3-1-1 symmetry, there is only a relevant scalar sextet among the six breaking the 3-3-1-1 symmetry which defines the right-handed neutrino mass scale and the matter parity. It is quoted as 
\be \sigma = \begin{pmatrix}\sigma^{++}_{11} & \fr{1}{\sqrt{2}}\sigma^{+}_{12} & \fr{1}{\sqrt{2}} \sigma^{+}_{13}\\
\fr{1}{\sqrt{2}} \sigma^{+}_{12} & \sigma^0_{22} & \fr{1}{\sqrt{2}} \sigma^0_{23} \\
\fr{1}{\sqrt{2}} \sigma^{+}_{13} & \fr{1}{\sqrt{2}} \sigma^0_{23} & \sigma^0_{33}
\end{pmatrix}\sim (1,6,2/3,-2/3),\ee where we relabel the original fields for brevity, and the coefficients $\fr{1}{\sqrt{2}}$ count for field's renormalization. The remaining scalar sextets do not affect the 3-3-1-1 symmetry and particle mass spectra, which might be manifestly neglected.\footnote{Indeed, they can otherwise be made heavy with a mass, by adjusting appropriate parameters, beyond $w$ scale, hence being integrated out.} Conversely, similar to the fermion representations as decomposed, the scalar multiplets $\eta$, $\rho$, $\varphi$, and $\sigma$ are assumed to survive at the 3-3-1-1 scale. Up to the trinification scale, they unite with necessary heavy particles to reveal a dark grand unification. The last regime is an assumption, while the physics in the 3-3-1-1 regime can be tested in experiments.  

\begin{table}[h]
\bc
\begin{tabular}{r|cccccccccccccccccccccccccc}
\hline\hline
Particle & $\nu$ & $e$ & $u$ & $d$ & $N$ & $E$ & $D$ & $G$ & $A_{3,8}$& $W$ & $X$ & $Y$ & $B$ & $C$ & $\eta_{1,2}$ & $\rho_{1,2}$ & $\varphi_3$ & $\eta_3$ & $\rho_3$ & $\varphi_{1,2}$ & $\sigma_{11,12,22}$ & $\sigma_{13,23}$ & $\sigma_{33}$\\
\hline 
$B-L$ & $-1$ & $-1$ & $\fr 1 3 $ & $\fr 1 3 $ & 0 & 0 & $-\fr{2}{3}$ & 0 & 0 & 0 & 1 & 1 & 0 & 0 & 0 & 0 & 0 & $-1$ & $-1$ & 1 & 0 & $-1$ & $-2$ \\
$M_P$ & $+$ & $+$ & $+$ & $+$ & $-$ & $-$ & $-$ & $+$ & $+$ & $+$ & $-$ & $-$ & $+$ & $+$ & $+$ & $+$ & $+$ & $-$ & $-$ & $-$ & $+$ & $-$ & $+$\\
\hline \hline
\end{tabular}
\caption[]{\label{tab3} $B-L$ number and matter parity of all particles in the 3-3-1-1 model.}
\ec
\end{table}
With the 3-3-1-1 quantum numbers for multiplets, we can derive the $B-L$ charge of all the 3-3-1-1 model particles, as collected in Table \ref{tab3}. By the way, we supply the matter parity, i.e. $M_P=(-1)^{3(B-L)+2s}$, which is detailedly derived below. Let us remind the reader that $(A_{3,8},W,X,Y)$ are the gauge bosons of $SU(3)_L$, i.e.
\be A = A_{n} T_{nL}=\begin{pmatrix}\fr 1 2 A_3+\fr{1}{2\sqrt{3}} A_8 & \fr{1}{\sqrt{2}}W^+ & \fr{1}{\sqrt{2}}X^+\\
\fr{1}{\sqrt{2}}W^- & -\fr 1 2 A_3+\fr{1}{2\sqrt{3}} A_8  & \fr{1}{\sqrt{2}}Y^0\\
\fr{1}{\sqrt{2}}X^- & \fr{1}{\sqrt{2}}Y^{0*}& -\fr{1}{\sqrt{3}} A_8  \end{pmatrix},\ee where $W^\pm=(A_1\mp i A_2)/\sqrt{2}$, $X^\pm=(A_4\mp i A_5)/\sqrt{2}$, and $Y^{0,0*}=(A_6\mp i A_7)/\sqrt{2}$. Note that the subscript $_L$ on the fields was removed, without confusion; whereas, the gauge bosons of $U(1)_{X,N}$ are denoted by $B,C$, respectively.  

Now that the scalar fields can develop VEVs, such as 
\be \langle \eta\rangle = 
\begin{pmatrix}
\fr{u}{\sqrt{2}}\\
0\\
0
\end{pmatrix},\hs
\langle \rho\rangle = 
\begin{pmatrix}
0\\
\fr{v}{\sqrt{2}}\\
0
\end{pmatrix},\hs 
\langle \varphi\rangle = 
\begin{pmatrix}
0\\
0\\
\fr{w}{\sqrt{2}}
\end{pmatrix},\hs 
\langle \sigma\rangle =\begin{pmatrix}
0 & 0 & 0\\
0 & \fr{\kappa}{\sqrt{2}} & 0\\
0 & 0 & \fr{\La}{\sqrt{2}} 
\end{pmatrix}. 
\ee It is noted that only scalar fields that are electrically neutral and even under the matter parity can have a VEV. In other words, the VEVs of $\rho^0_3$, $\varphi^0_2$, and $\sigma^0_{23}$ vanish since they are odd under the matter parity. Additionally, $w$ and $\La$ break the 3-3-1-1 symmetry, giving mass for new particles, while $u$ and $v$ break the standard model symmetry, supplying mass for usual particles. To be consistent with the standard model, we assume $w,\La\gg u,v$. Additionally, $\kappa$ is the VEV of a scalar triplet under the standard model symmetry contributing to the $\rho$-parameter. The current constraint for $\rho$-parameter deviation requires $\kappa\lesssim \mathcal{O}(1) $ GeV. 

The scheme of the 3-3-1-1 symmetry breaking is given by
\bc \begin{tabular}{c}
$SU(3)_C\otimes SU(3)_L\otimes U(1)_X\otimes U(1)_N$ \\
$\downarrow w,\La$\\
$SU(3)_C\otimes SU(2)_L\otimes U(1)_Y\otimes M_P$\\
$\downarrow u,v$\\
$SU(3)_C\otimes U(1)_Q\otimes M_P$. 
\end{tabular}\ec   
The charge $B-L=\fr{2}{\sqrt{3}} T_{8L}+N$ is the residual symmetry of $SU(3)_L\otimes U(1)_N$ because it annihilates the vacua $[B-L]\langle \eta\rangle = 0$, $[B-L]\langle \rho\rangle =0$, and $[B-L]\langle \varphi\rangle =0$ for $u,v,w\neq 0$. It transforms a field as $f\to f'=U(\al)f$ with $U(\al)=e^{i\al(B-L)}$, where $\al$ is a transforming parameter. Next, $B-L$ is broken by $\sigma$ due to $\{B-L,\langle \sigma\rangle \}= \mathrm{diag}(0,0,-\sqrt{2}\La)\neq 0$. However, it can have a remnant conserving such vacuum, i.e. $U(\al)\langle \sigma\rangle U^T(\al)=\langle \sigma\rangle$. This gives $e^{-2i\al}=1$, thus $\al=k\pi$ for $k$ integer. We deduce $M_P=U(k\pi)=e^{i k\pi (B-L)}$. Looking at Tab. \ref{tab3}, we find that $M_P=1$ for every field when the minimum of $|k|$ is $|k|=6$, except for the identity $k=0$. Thus, we write $M_P=\{1,p,p^2,p^3,p^4,p^5\}$, where $p=e^{i\pi (B-L)}$ and $p^6=1$. $M_P$ is a $Z_6$ symmetry factorized to $M_P=Z_2\otimes Z_3$, where $Z_2=\{1,p^3\}$ is the invariant subgroup of $M_P$, while the quotient group of $M_P$ by $Z_2$ is $Z_3=M_P/Z_2=\{[1],[p^2],[p^4]\}$, where each coset element has the form $[x]=xZ_2=\{x,xp^3\}$, thus $[1]=[p^3]=Z_2$, $[p]=[p^4]=\{p,p^4\}$, and $[p^2]=[p^5]=\{p^2,p^5\}$. That said, $Z_2$ and $Z_3$ are generated by the generators $p^3=(-1)^{3(B-L)}$ and $[p^2]=[\om^{3(B-L)}]$, respectively, where $\om=e^{i2\pi/3}$ is cube root of unity. The group $Z_3$ transforms nontrivially only for quarks, i.e. $[p^2]=[w]\to w$ for $u,d,D$, which is isomorphic to the center of the color group. That said, $Z_3$ is accidentally conserved by the color group. Hence, omitting $Z_3$, the residual symmetry $M_P$ is only $Z_2$. We can identify $M_P=p^3$ and multiply it by the spin parity $S_P=(-1)^{2s}$ as conserved by the Lorentz symmetry, i.e.
\be M_P=p^3\times S_P=(-1)^{3(B-L)+2s}.\ee This result is similar to that in a canonical seesaw \cite{dongmp}.       

Although the trinification is broken, the unification of dark matter and normal matter is still partly seen at the 3-3-1-1 level, through the matter parity of particle multiplets, 
\bea &&Q = \begin{pmatrix} +\\ +\\ - \end{pmatrix},\hs \psi = \begin{pmatrix} +\\ +\\ - \end{pmatrix},\hs \psi_{\nu} = \begin{pmatrix} - \\ 
- \\ 
+  \end{pmatrix},\hs \psi_{e} = \begin{pmatrix} - \\ 
- \\ 
+  \end{pmatrix},\\ 
&& \eta = \begin{pmatrix} +\\ +\\ - \end{pmatrix},\hs \rho = \begin{pmatrix} +\\ +\\ - \end{pmatrix},\hs \varphi = \begin{pmatrix} - \\ 
- \\ 
+  \end{pmatrix},\hs \sigma = \begin{pmatrix} + & + & -\\ 
+ & + & -\\ 
- & -& + \end{pmatrix},\hs A = \begin{pmatrix} + & + & -\\ 
+ & + & -\\ 
- & -& + \end{pmatrix}, \eea where the left and right chirality notations are conveniently suppressed, without confusion, since they have the same matter parity.   

Remarks are given in order
\ben
\item The unwanted scalar vacua associated with $\rho^0_3$, $\varphi^0_2$, and $\sigma^0_{23}$ are suppressed by the matter parity conservation, which cannot do it in the corresponding 3-3-1 version. 
\item The unwanted Yukawa interactions which lead to even (normal) and odd (exotic) fermion mixing are suppressed by the matter parity conservation, which cannot prevent it in the relevant 3-3-1 version.
\item The lightest of matter parity odd fields, say $N_{L,R}$, $E^0_{L,R}$, $\rho^0_3$, $\varphi^0_2$, $\sigma^0_{23}$, and $Y^0$, is stabilized by the matter parity conservation, responsible for dark matter. It is clear that the number of scalar and vector dark matter candidates are significantly reduced, compared with the previous model.   
\een

\section{\label{pheno} Phenomenology of the universal 3-3-1-1 model}

\subsection{Fermion mass}

Yukawa Lagrangian is given by
\bea \mathcal{L}_{\mathrm{Yuk}} &=& h^u \bar{Q}_L \eta u_R+h^d \bar{Q}_L \rho d_R + h^D\bar{Q}_L\varphi D_R\crn
&&+h^e \bar{\psi}_L \rho \psi^c_{eL} + h^\nu \bar{\psi}_L \eta\psi^c_{\nu L} + h^E \bar{\psi}_{eL} \varphi \psi^c_{\nu L} + x \bar{\psi}^c_{\nu L} \sigma \psi_{\nu L} \crn
&& + h^N \bar{\psi}_L \varphi^* N_R + y \bar{\psi}_{eL} \eta^* N_R + z \bar{\psi}_{\nu L} \rho^* N_R + \mu N_R N_R+H.c.,
 \eea  where family indices are suppressed, which can easily be generalized. 
 
 When the scalar fields develop VEVs, as given, the fermions obtain a mass. The charged leptons and quarks gain masses, as usual,
 \bea && m_e=-h^e\fr{v}{\sqrt{2}},\hs m_{E^-}=-h^E\fr{w}{\sqrt{2}},\\ 
 && m_u = -h^u \fr{u}{\sqrt{2}},\hs m_d = -h^d \fr{v}{\sqrt{2}},\hs m_D = -h^D \fr{w}{\sqrt{2}}.\eea  
 
 Neutrinos possess a mass Lagrangian
 \be \mathcal{L}_{\mathrm{Yuk}}\supset -\fr 1 2 \begin{pmatrix} 
 \bar{\nu}_L & \bar{\nu}^c_L \end{pmatrix} 
 \begin{pmatrix}
 0 & m \\
 m & M \end{pmatrix}
 \begin{pmatrix}
 \nu^c_R\\
 \nu_R
 \end{pmatrix}+H.c.,\label{ed42}\ee where $m=h^\nu \fr{u}{\sqrt{2}}$ is the Dirac mass, while $M=-\sqrt{2} x \La$ is the right-handed Majorana mass. Notice that $\nu^c_L=(\nu_R)^c$ and $\nu^c_R=(\nu_L)^c$, where the superscript $^c$ indicates charge conjugation. Because of $u\ll \La$, the active neutrino $\sim \nu_L$ gains an appropriate small mass via seesaw, 
 \be m_\nu \simeq  -\fr{m^2}{M}=\fr{(h^\nu)^2u^2}{2\sqrt{2}x\La},\label{ed43}\ee  while the sterile neutrino $\sim \nu_R$ has a large mass approximate to $M$.     
 
Neutral fermions $(E^0_{L,R},N_{L,R})$ receive a mass Lagrangian,
\be \mathcal{L}_{\mathrm{Yuk}}\supset -\fr 1 2 \begin{pmatrix}
\bar{N}_L & \bar{N}^c_L & \bar{E}^0_L & \bar{E}^{0c}_L\end{pmatrix}
\begin{pmatrix}
0 & m_N & h^e\fr{v}{\sqrt{2}} & -h^\nu \fr{u}{\sqrt{2}}\\
m_N & -2\mu & -y\fr{u}{\sqrt{2}} & -z\fr{v}{\sqrt{2}}\\
h^e\fr{v}{\sqrt{2}} & -y\fr{u}{\sqrt{2}} & 0 & m_{E^0}\\
-h^\nu \fr{u}{\sqrt{2}} & -z\fr{v}{\sqrt{2}} & m_{E^0} & -\sqrt{2}x\kappa
\end{pmatrix}
\begin{pmatrix}
N^c_R \\ N_R \\ E^{0c}_R \\ E^{0}_R\end{pmatrix}, \label{ed44}
\ee  where $m_N=-h^N\fr{w}{\sqrt{2}}$ and $m_{E^0}=h^E\fr{w}{\sqrt{2}}$ are Dirac masses of $N$ and $E^0$, respectively. Because of $w\gg u,v\gg \mu,\kappa$, the Dirac masses $m_{N}$ and $m_{E^0}$ are biggest. Hence, this mass matrix would yield physical pseudo-Dirac states, $N_L$ vs. $N_R$, as well as $E^0_L$ vs. $E^0_R$. 

Indeed, at the leading order, the mass matrix has the form,
\be \begin{pmatrix}
0 & m_N & 0 & 0\\
m_N & 0 & 0 & 0\\
0 & 0 & 0 & m_{E^0}\\
0 & 0 & m_{E^0} & 0
\end{pmatrix},\label{ed45}
\ee which is diagonalized into $\mathrm{diag}(m_N,-m_N,m_{E^0},-m_{E^0})$ by the following orthogonal matrix that relates to the new basis,
\be \begin{pmatrix}
N^c_R \\ N_R \\ E^{0c}_R \\ E^{0}_R\end{pmatrix} = \begin{pmatrix} \fr{1}{\sqrt{2}} & \fr{1}{\sqrt{2}} & 0 & 0 \\
\fr{1}{\sqrt{2}} & -\fr{1}{\sqrt{2}} & 0 & 0\\
0 & 0 & \fr{1}{\sqrt{2}} & \fr{1}{\sqrt{2}}\\
0& 0& \fr{1}{\sqrt{2}} & -\fr{1}{\sqrt{2}}\end{pmatrix} \begin{pmatrix}
N_{1R} \\ N_{2R} \\ E_{1R} \\ E_{2R}\end{pmatrix}.\label{ed46}\ee At the leading order, $N_1$ and $N_2$ have degenerate masses $m_N$ and $-m_D$, as well as $E_1$ and $E_2$ do so with $m_{E^0}$ and $-m_{E^0}$. Assuming that $\mu\sim \kappa \sim (u,v)^2/\La$, the approximation at the next to leading order yields mass eigenvalues,
\be  m_{N_1}\simeq m_N-\mu,\hs m_{N_2}\simeq - m_N-\mu,\hs m_{E_1}\simeq m_{E^0}-\fr{x\kappa}{\sqrt{2}},\hs m_{E_2}\simeq -m_{E^0}-\fr{x\kappa}{\sqrt{2}}. \label{ed47} \ee This means that we have two pairs of pseudo-Dirac states $(N_1,N_2)$ and $(E_1,E_2)$ with separated masses $|m_{N_1}|\neq |m_{N_2}|$ and $|m_{E_1}|\neq |m_{E_2}|$, respectively. As it is seen, the $E^0_{L,R}$ mass splitting is necessary for one of them to be realistic dark matter candidate.

\subsection{Radiative correction to neutrino mass}

\begin{figure}[h]
\bc
\includegraphics[]{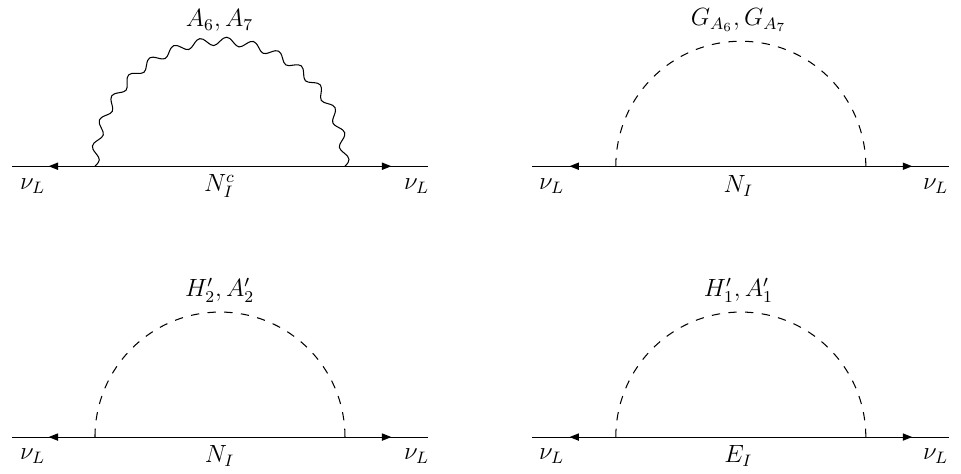}
\caption[]{\label{fig6} Radiative corrections of the dark gauge/Higgs fields to the neutrino mass.}
\ec
\end{figure} 

The neutrino mass in (\ref{ed43}) is given by a canonical seesaw, which results from the mixing of the active left-handed neutrino $\nu_L$ with the Majorana right-handed neutrino $\nu_R$ as in (\ref{ed42}). Further, there is an alternative source contributing to neutrino mass, which is radiatively induced by dark/odd fields, say $(Y^0,N_L)$, $(\varphi^0_2, N_R)$, and $(\rho^0_3,E^0_L)$, which couple to $\nu_L$. It is noted that the tree-level canonical seesaw also furnishes radiative corrections to neutrino mass, mediated by the normal/even fields: the Higgs/Goldstone $H,G_Z$ that couple $\nu_L$ to $\nu_R$ and the gauge $Z$ that couples two $\nu_L$'s, but these contributions are negligible, as shown in the appendix of \cite{dongloidn}. That said, the dominant radiative contributions are described by Feynman diagrams in Fig. \ref{fig6}, whose intermediate fields are given in the basis of mass eigenstates, i.e. $Y^0=(A_6-iA_7)/\sqrt{2}$, $\varphi^0_2 = s_\al (G_{A_7}+iG_{A_6})/\sqrt{2}+c_\al (H'_2+iA'_2)/\sqrt{2}$, with $t_\al=w/\sqrt{2}\La$, $\rho^0_3 = (H'_1+iA'_1)/\sqrt{2}$, and $N_{L},N_R$ ($E^0_{L},E^0_R$) are related to $N_I$ ($E_I$) for $I=1,2$.\footnote{In generic $R_\xi$-gauge, the gauge boson diagrams mediated by $A_{6,7}$ must include the relevant Goldstone diagrams by $G_{A_{6,7}}$, which differ from the usual scotogenic ones mediated by $H'_{2},A'_{2}$ and $H'_{1},A'_{1}$.} Notice that $N_I$ and $E_I$ are pseudo- (exactly, quasi-) Dirac states and substantially separated, such as $(N^c,N)^T=U(N_1,N_2)^T$ and $(E^{0c},E^0)^T=U(E_1,E_2)^T$, in which \be U=\begin{pmatrix}\fr{1}{\sqrt{2}}& \fr{1}{\sqrt{2}}\\
\fr{1}{\sqrt{2}}&-\fr{1}{\sqrt{2}}\end{pmatrix},\ee which can be extracted from (\ref{ed46}). This leads to useful relations, $m_N=U^*_{1I}U^*_{2I}m_{N_I}$ and $0=U^*_{1I}U^*_{1I}m_{N_I}$, where the indices $I$ are summed, which imply $m_N U_{2I}=U^*_{1I}m_{N_I}$ (no sum over $I$). Further, the Goldstone bosons $G_{A_7}\equiv S'_2$, $G_{A_6}\equiv  P'_2$ and the Higgs bosons $H'_{1,2}$, $A'_{1,2}$ as well as their masses are determined in the subsequent sections.\footnote{The identification of the Goldstone bosons, $G_{A_6}$ as a pseudo-scalar $P'_2$ and $G_{A_7}$ as a scalar, can be verified via a gauge rotation to the unitarity gauge.} Since our theory is renormalizable, as well as the neutrino mass vanishes at the tree level as the 11 element in~(\ref{ed42}), every radiative correction to the neutrino mass must be finite.  

Applying the Feynman rules in generic $R_\xi$-gauge, the radiative neutrino mass is derived in the form of $\mathcal{L}\supset -\fr 1 2 \bar{\nu}_{L} m^{\mathrm{rad}}_\nu \nu^c_{R}+H.c.$, where $m^{\mathrm{rad}}_\nu=(m^{\mathrm{rad}}_\nu)_{\mathrm{G}}+(m^{\mathrm{rad}}_\nu)_{\mathrm{H}}$, in which 
\bea -i (m^{\mathrm{rad}}_\nu)_{\mathrm{G}} P_R &=& \int \fr{d^4p}{(2\pi)^4}\left(\fr{-ig}{2}\ga^\mu P_L U^*_{1I}\right)\fr{i}{-p\!\!\!/-m_{N_I}}\left(\fr{ig}{2}\ga^\nu P_R U^*_{1 I}\right)\crn
&&\times \fr{-i }{p^2-m^2_{A_6}}\left(g_{\mu\nu}-\fr{(1-\xi)p_\mu p_\nu}{p^2-\xi m^2_{A_6}}\right) \crn
&&+ \int \fr{d^4p}{(2\pi)^4}\left(\fr{-g}{2}\ga^\mu P_L U^*_{1I}\right)\fr{i}{-p\!\!\!/-m_{N_I}}\left(\fr{g}{2}\ga^\nu P_R U^*_{1 I}\right)\crn
&&\times \fr{-i }{p^2-m^2_{A_7}}\left(g_{\mu\nu}-\fr{(1-\xi)p_\mu p_\nu}{p^2-\xi m^2_{A_7}}\right) 
\crn
&&+ \int \fr{d^4 p}{(2\pi)^4}\left(\fr{gm_N}{2m_{Y}}P_R U_{2I}\right)\fr{i}{p\!\!\!/  - m_{N_I}}\left(\fr{gm_N}{2m_{Y}}P_R U_{2I}\right)\fr{i}{p^2-\xi m^2_{A_6}}
\crn
&&+\int \fr{d^4 p}{(2\pi)^4}\left(\fr{igm_N}{2m_{Y}} P_R U_{2 I}\right)\fr{i}{p\!\!\!/  - m_{N_I}}\left(\fr{igm_N}{2m_{Y}}P_R U_{2 I}\right)\fr{i}{p^2-\xi m^2_{A_7}},\eea coming from the contributions of the gauge/Goldstone fields $A_{6,7}$ and $G_{A_{6,7}}$, while 
\bea 
-i (m^{\mathrm{rad}}_\nu)_{\mathrm{H}} P_R &=& \int \fr{d^4 p}{(2\pi)^4}\left(\fr{-ih^Nc_\al}{\sqrt{2}} P_R U_{2 I}\right)\fr{i}{p\!\!\!/  - m_{N_I}}\left(\fr{-ih^Nc_\al}{\sqrt{2}} P_R U_{2 I}\right)\fr{i}{p^2- m^2_{H'_2}}\crn
&&+ \int \fr{d^4 p}{(2\pi)^4}\left(\fr{-h^Nc_\al}{\sqrt{2}} P_R U_{2 I}\right)\fr{i}{p\!\!\!/  - m_{N_I}}\left(\fr{-h^Nc_\al}{\sqrt{2}} P_R U_{2 I}\right)\fr{i}{p^2- m^2_{A'_2}}\crn
&&+\int \fr{d^4 p}{(2\pi)^4}\left(\fr{-ih^e}{\sqrt{2}} P_R U_{1 I}\right)\fr{i}{p\!\!\!/  - m_{E_I}}\left(\fr{-ih^e}{\sqrt{2}} P_R U_{1 I}\right)\fr{i}{p^2- m^2_{H'_1}}\crn
&&+ \int \fr{d^4 p}{(2\pi)^4}\left(\fr{h^e}{\sqrt{2}} P_R U_{1 I}\right)\fr{i}{p\!\!\!/  - m_{E_I}}\left(\fr{h^e}{\sqrt{2}} P_R U_{1 I}\right)\fr{i}{p^2- m^2_{A'_1}},
\eea arising from the contributions of the Higgs fields $H'_{1,2}$ and $A'_{1,2}$. Let us note that $m_Y=(g/2)\sqrt{w^2+2\La^2}\simeq m_{A_7}\simeq m_{A_8}$. Additionally, the propagator of the charge conjugated field $N^c$ takes a form $i/(-p\!\!\!/-m_{N_I})$ \cite{dehk}. We estimate further,
\bea  (m^{\mathrm{rad}}_\nu)_{\mathrm{G}}  &=& i\int \fr{d^4 p }{(2\pi)^4}\fr{3g^2U^*_{1I}U^*_{1I}m^3_{N_I}}{4m^2_Y}\fr{m^2_{A_{6}}-m^2_{A_7}}{(p^2-m^2_{N_I})(p^2-m^2_{A_6})(p^2-m^2_{A_7})},\label{radg}\\
(m^{\mathrm{rad}}_\nu)_{\mathrm{H}}  &=& i\int \fr{d^4 p }{(2\pi)^4}\fr{h^N h^N c^2_\al U_{2I}U_{2I} m_{N_I}}{2}\fr{m^2_{H'_2}-m^2_{A'_2}}{(p^2-m^2_{N_I})(p^2-m^2_{H'_2})(p^2-m^2_{A'_2})}\crn
&&+ i\int \fr{d^4 p }{(2\pi)^4}\fr{h^e h^e U_{1I}U_{1I} m_{E_I}}{2}\fr{m^2_{H'_1}-m^2_{A'_1}}{(p^2-m^2_{E_I})(p^2-m^2_{H'_1})(p^2-m^2_{A'_1})}.\label{radh}\eea    

\begin{figure}[h]
\bc
\includegraphics[]{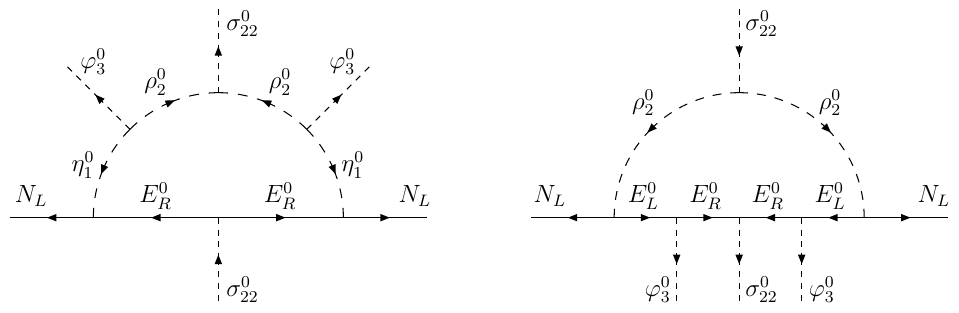}
\caption[]{\label{fig7} Radiative contribution to Majorana $N_L$ mass.}
\ec
\end{figure}

With the aid of quasi-Dirac approximations as given in (\ref{ed46}) and (\ref{ed47}), the radiative neutrino mass (\ref{radg}) becomes
\be (m^{\mathrm{rad}}_\nu)_{\mathrm{G}} = -\fr{3ig^2m^2_N\mu}{4m^2_Y}\int \fr{d^4p}{(2\pi)^4}\fr{(3p^2-m^2_N)(m^2_{A_{6}}-m^2_{A_7})}{(p^2-m^2_{N})^2(p^2-m^2_{A_6})(p^2-m^2_{A_7})}.\ee With the aid of the formula,\bea I &=&\int \fr{d^4p}{(2\pi)^4}\fr{(\al p^2+\beta c)(a-b)}{(p^2-c)^2(p^2-a)(p^2-b)}\crn
&=&-\fr{i}{16\pi^2}\left\{\al \left(\fr{a\ln \fr{a}{c}}{a-c}-\fr{b\ln\fr{b}{c}}{b-c}\right)+(\al+\beta)c\left[\fr{a-b}{(a-c)(b-c)}+\fr{a\ln \fr{a}{c}}{(a-c)^2}-\fr{b\ln\fr{b}{c}}{(b-c)^2}\right]\right\}\crn
&\simeq&  -\fr{i}{16\pi^2}\fr{a-b}{\bar{x}-c}\left\{\fr{\al \bar{x}+(\al+2\beta)c}{\bar{x}-c}-\fr{c[(2\al+\beta)\bar{x}+\beta c]}{(\bar{x}-c)^2}\ln\fr{\bar{x}}{c} \right\},\label{usuf}\eea for $b\simeq a$ and $\bar{x}=(a+b)/2$, we have further,
\be (m^{\mathrm{rad}}_\nu)_{\mathrm{G}} = -\fr{3g^4}{32\pi^2}\fr{\mu \kappa}{\La}\fr{m^2_N}{m^2_Y}\fr{\La^2}{m^2_Y-m^2_N}\left[\fr{3m^2_Y+m^2_N}{m^2_Y-m^2_N}-\fr{m^2_N(5m^2_Y-m^2_N)}{(m^2_Y-m^2_N)^2}\ln \fr{m^2_Y}{m^2_N}\right],\label{kqn2} \ee with noting that $m^2_{A_6}-m^2_{A_7}=2g^2\kappa \La$. Since $m_N\sim m_Y\sim \La$ and $\kappa/\La\sim 10^{-3}$, we estimate $(m^{\mathrm{rad}}_\nu)_{\mathrm{G}}\sim 10^{-6}\mu$. Appropriate to the measured neutrino mass requires $\mu \lesssim 0.1 $ MeV. A question raised is that this small value is maintained against radiative corrections. The most dangerous contribution comes from a coupling of $N_L$ with $E^0_L$ via $h^e$ coupling, specially for the third family of tau. Since both $N_L$ and $E^0_L$ are odd fields, they couple only to an even scalar field, such as $\rho^0_2$, given by $-h^e\bar{N}_L\rho^0_2 (E^{0}_L)^c$. Notice that $E^0_L$ couples to $E^0_R$ via $\varphi^0_3$ through $h^E$ coupling. And, $E^0_R$ by itself couples to $\sigma^0_{22}$ via $x$ coupling. Hence, at the tree level, the Majorana mass of $N_L$ can be induced by an inverse seesaw [cf. (\ref{ed44})] \cite{invs1,invs2,invs3} to be $\mu_{\mathrm{eff}}\sim (m_e/m_{E^0})^2 x\kappa \lesssim 10^{-6} x \kappa \ll 0.1$ MeV, with a $\kappa$ at GeV, even for the third family of tau, i.e. $m_\tau= - h^\tau v/\sqrt{2}\simeq 1.777$ GeV. At the one loop level, the Majorana mass of $N_L$ comes from the right diagram as depicted in Fig. \ref{fig7}, where the contribution of the left diagram to this mass arises from a coupling of $N_L$ with $E^0_R$ (i.e. $h^\nu$) as also included for completeness, where both diagrams are given in flavor basis, for brevity. Notice that $E^0_{L,R}$ is pseudo-Dirac with mass splitting proportional to $x\kappa$. The right diagram contribution can conventionally be written as \be \mu_{\mathrm{eff}}\simeq -\fr{(h^e)^2}{16\sqrt{2}\pi^2}x\kappa \fr{m^2_{R_2}-m^2_{I_2}}{m^2_{\rho^0_2}-m^2_{E^0}}\left[\fr{m^2_{\rho^0_2}+3m^2_{E^0}}{m^2_{\rho^0_2}-m^2_{E^0}}-\fr{m^2_{E^0}(3m^2_{\rho^0_2}+m^2_{E^0})}{m^2_{\rho^0_2}-m^2_{E^0}}\ln\fr{m^2_{\rho^0_2}}{m^2_{E^0}}\right],\ee where we denote $m^2_{\rho^0_2}=(m^2_{R_2}+m^2_{I_2})/2$ and note that $R_2,I_2$ stand for physical real and imaginary fields associated with $\rho^0_2$, respectively, for which $m^2_{R_2,I_2}$ are some physical masses of $H,H_1,\mathcal{A}_{1,3}$ in mass basis. Since $m_{R_2}\sim m_{I_2}\sim m_{E^0}$, the one-loop correction is proportional to $\mu_{\mathrm{eff}}\sim (1/16\pi^2) (h^e)^2 x \kappa \lesssim 0.1$ MeV, even for $h^\tau\sim 0.1$ (i.e. $u/v\sim 10$), as expected. Like the evaluation for the right diagram, the left diagram contribution is proportional to $(1/16\pi^2)(h^\nu)^2 x \kappa$, which is radically smaller than the right diagram contribution, thus 0.1 MeV, since $h^\nu$ would be as small as that of electron coupling. Last, but not least, the radiative Majorana masses of $N_R$ (if $\mu=0$) and $E^0_L$ must be suppressed by the quasi-Dirac approximation for $E^0_{L,R}$, i.e. $x\kappa$, and the relevant couplings $y,z,h^e$, similar to that of $N_L$. Given that $y,z$ are as small as those of charged leptons, these radiative Majorana masses are small (i.e., safe), like that of $N_L$, as desirable.     

Similarly, with the help of quasi-Dirac approximations as in (\ref{ed46}) and (\ref{ed47}), the radiative neutrino mass (\ref{radh}) takes the form,
\bea (m^{\mathrm{rad}}_\nu)_{\mathrm{H}}  &=& -\fr{i(h^N c_\al)^2\mu}{2}\int \fr{d^4 p }{(2\pi)^4}\fr{(p^2+m^2_N)(m^2_{H'_2}-m^2_{A'_2})}{(p^2-m^2_{N})^2(p^2-m^2_{H'_2})(p^2-m^2_{A'_2})}\crn
&&- \fr{i(h^e)^2x\kappa}{2\sqrt{2}}\int \fr{d^4 p }{(2\pi)^4}\fr{(p^2+m^2_{E^0})(m^2_{H'_1}-m^2_{A'_1})}{(p^2-m^2_{E^0})^2(p^2-m^2_{H'_1})(p^2-m^2_{A'_1})}.\eea  With the help of (\ref{usuf}), as well as the mass splittings $m^2_{H'_2}-m^2_{A'_2}\simeq 2s^2_\al (\la_{11} -\la_{16})\kappa \La$ and $m^2_{H'_1}-m^2_{A'_1}\simeq 2\sqrt{2} f_2 \La$, we have further
\bea (m^{\mathrm{rad}}_\nu)_{\mathrm{H}}  &=& -\fr{(h^N s_\al c_\al)^2}{16\pi^2}\fr{\mu\kappa}{\La} \fr{(\la_{11}-\la_{16})\La^2}{m^2_{H'^0_2}-m^2_N}\left[\fr{m^2_{H'^0_2}+3m^2_N}{m^2_{H'^0_2}-m^2_N}-\fr{m^2_N(3m^2_{H'^0_2}+m^2_N)}{(m^2_{H'^0_2}-m^2_N)^2}\ln\fr{m^2_{H'^0_2}}{m^2_N}\right]\crn
&&- \fr{(h^e)^2}{16\pi^2} x\kappa \fr{f_2\La}{m^2_{H'^0_1}-m^2_{E^0}}\left[\fr{m^2_{H'^0_1}+3m^2_{E^0}}{m^2_{H'^0_1}-m^2_{E^0}}-\fr{m^2_{E^0}(3m^2_{H'^0_1}+m^2_{E^0})}{m^2_{H'^0_1}-m^2_{E^0}}\ln\fr{m^2_{H'^0_1}}{m^2_{E^0}}\right].\label{kqn3}\eea  The first term contribution due to the dark doublet Higgs $H'^0_2=(H'_2+i A'_2)/\sqrt{2}$ where $m^2_{H'^0_2}=(m^2_{H'_2}+m^2_{A'_2})/2$ is similar to the above gauge contribution, given that $h^N\sim t_\al \sim 1$ (i.e., $s_\al\sim c_\al \sim 1/\sqrt{2}$), which requires $\mu\lesssim 0.1$ MeV. We take $u\sim v$ for the following discussion, without loss of generality. The second term contribution arising from the dark singlet Higgs $H'^0_1=(H'_1+iA'_1)/\sqrt{2}$ with $m^2_{H'^0_1}=(m^2_{H'_1}+m^2_{A'_1})/2$ is negligible for the first family coupling of lepton, say $h^e\sim 10^{-6}$, but is considerable according to the remaining lepton families. This contribution is safely below the measured neutrino mass for the second family coupling of lepton, say $h^\mu\sim 10^{-3}$, given that $x\kappa\lesssim 0.015$ GeV. This leads to $x\lesssim 0.015$ for $\kappa\sim $ GeV. While the first family right-handed neutrino is possibly beyond TeV, the second family right-handed neutrino gains a Majorana mass $M=-\sqrt{2}x\La \sim 100$ GeV, which is at the weak scale, for a benchmark $\La\sim 5$ TeV. However, for the third family of lepton, i.e. $h^\tau\sim 0.01$, it acquires $x\kappa\lesssim 1.5\times 10^{-4}$ GeV, which demands that $x\lesssim 1.5\times 10^{-4}$, since $\kappa\sim$ GeV. It follows that the third family right-handed neutrino obtains a Majorana mass at $M=-\sqrt{2}x\La\sim 1$~GeV, which is significantly below the weak scale. In this case, the tree-level seesaw works if the corresponding Dirac neutrino coupling is $h^\nu\sim 10^{-7}$, below that of electron, while for the previous cases of the first and second families, $h^\nu$'s are proportional to that of electron, i.e. $h^\nu\sim 10^{-6}$, as usual. In summary, including the radiative corrections to neutrino mass, the neutrino mass matrix (\ref{ed42}) now takes the form,
\be \mathcal{M}=\begin{pmatrix} m_\nu^{\mathrm{rad}} & m \\
m & M\end{pmatrix},\ee which is composed of type I and II seesaws. It yields an effective neutrino mass, by virtue of seesaw diagonalization, such as 
\be m^{\mathrm{eff}}_\nu\simeq -m M^{-1}m^T+m^{\mathrm{rad}}_\nu,\ee which summarizes those of (\ref{ed43}), (\ref{kqn2}), and (\ref{kqn3}), and this form is generalized for three families. In short, the model predicts, for $u\sim v$, a right-handed neutrino with Majorana mass at 1 GeV, one other right-handed neutrino with Majorana mass at the weak scale, while the remaining right-handed neutrino may be heavy, beyond TeV. Given that $u/v\sim 10$, the masses of the two light right-handed neutrinos are reduced down to two order of magnitudes, i.e. 10 MeV and 1 GeV, while the mass of the heaviest right-handed neutrino is retained.

\subsection{Gauge boson mass}

The gauge bosons obtain mass through 
\be \mathcal{L}_{\mathrm{kin}} \supset \sum_{\Phi =\eta,\rho,\varphi,\sigma} (D^\mu \Phi)^\dagger (D_\mu \Phi), \ee when the scalar fields develop VEVs. The covariant derivative is generally given by 
\be D_\mu = \pa_\mu + i g_s G_{n\mu}t_n + i g A_{n\mu}T_{nL}+ i g_X B_\mu X + i g_N C_\mu N,\ee where $(g_s,g,g_X,g_N)$, $(G_n, A_n, B,C)$, and $(t_n,T_{nL},X,N)$ are coupling constants, gauge bosons, and generators associated with the 3-3-1-1 groups, respectively.

The charged gauge bosons $W^\pm,X^\pm$ are physical fields by themselves with masses,
\bea && m^2_W=\fr{g^2}{4}(u^2+v^2+2\kappa^2)\simeq \fr{g^2}{4}(u^2+v^2),\\ && m^2_X=\fr{g^2}{4}(u^2+w^2+2\La^2)\simeq \fr{g^2}{4}(w^2+2\La^2).\eea
By contrast, the real and imaginary parts of non-Hermitian gauge boson $Y^{0,0*}=(A_6\mp i A_7)/\sqrt{2}$ are separated in mass, such as 
\bea  
 && m^2_{A_6}=\fr{g^2}{4}\left[v^2+w^2+2(\La+\kappa)^2\right] \simeq \fr{g^2}{4}(w^2+2\La^2),\\
 && m^2_{A_7}=\fr{g^2}{4}\left[v^2+w^2+2(\La-\kappa)^2\right] \simeq \fr{g^2}{4}(w^2+2\La^2).
 \eea In other words, the real and imaginary parts $A_6$ and $A_{7}$ are physical fields with respective masses, as given. This mass splitting is necessary for one of them, $A_6$ or $A_7$, to be realistic dark matter candidate, opposite to the old 3-3-1-1 model \cite{d1}. 
 
The neutral gauge bosons obtain a mass Lagrangian,
\be \mathcal{L}_{\mathrm{kin}}\supset \fr 1 2 
\begin{pmatrix} A_3 & A_8 & B & C\end{pmatrix} M^2
\begin{pmatrix} A_3 \\ A_8 \\ B \\ C\end{pmatrix},\ee where the mass matrix $M^2$ is derived by
\be \begin{pmatrix} 
\fr{g^2(u^2+v^2+4\kappa^2)}{4} &\fr{g^2(u^2-v^2-4\kappa^2)}{4\sqrt{3}} & \fr{-g^2 t_X( 2 u^2 +v^2+4\kappa^2)}{6}& \fr{g^2 t_N(-u^2+v^2+4\kappa^2)}{6}\\
\fr{g^2(u^2-v^2-4\kappa^2)}{4\sqrt{3}} & \fr{g^2(u^2+v^2+4\kappa^2 +4w^2+16\La^2)}{12} & \fr{g^2 t_X(-2u^2+v^2+4\kappa^2-2w^2-8\La^2)}{6\sqrt{3}} & \fr{g^2 t_N(-u^2-v^2-4\kappa^2-4w^2+8\La^2)}{6\sqrt{3}}\\
\fr{-g^2 t_X( 2 u^2 +v^2+4\kappa^2)}{6} & \fr{g^2 t_X(-2u^2+v^2+4\kappa^2-2w^2-8\La^2)}{6\sqrt{3}} & \fr{g^2t^2_X(4u^2+v^2+4\kappa^2+w^2+4\La^2)}{9} & \fr{g^2 t_X t_N (2u^2-v^2-4\kappa^2+2w^2-4\La^2)}{9}\\
\fr{g^2 t_N(-u^2+v^2+4\kappa^2)}{6} & \fr{g^2 t_N(-u^2-v^2-4\kappa^2-4w^2+8\La^2)}{6\sqrt{3}} & \fr{g^2 t_X t_N (2u^2-v^2-4\kappa^2+2w^2-4\La^2)}{9} & \fr{g^2 t^2_N (u^2+v^2+4\kappa^2+4w^2+4\La^2)}{9} 
\end{pmatrix},\ee where $t_X=g_X/g$ and $t_N=g_N/g$.  

From $Q=T_{3L}+\fr{1}{\sqrt{3}}T_{8L}+X$, we find the photon field, called $A$ without confusion, as
\be \fr{A}{e}=\fr{A_3}{g}+\fr{A_8}{\sqrt{3}g}+\fr{B}{g_X},\label{photf}\ee which is achieved by replacing each generator by its field per coupling, i.e. $Q$ by $A/e$, $T_{3L}$ by $A_3/g$, $T_{8L}$ by $A_8/g$, and $X$ by $B/g_X$, in the electric charge expression \cite{donglong}. Conversely, it is easily verified that $A$ is the eigenstate of $M^2$ corresponding to zero eigenvalue (photon mass) \cite{donglong}. Since the fields in (\ref{photf}) are normalized, we have 
\be \fr{1}{e^2}=\fr{1}{g^2}+\fr{1}{3g^2}+\fr{1}{g^2_X}.\ee Since $SU(3)_L\supset SU(2)_L$, the coupling $g$ of $SU(3)_L$ matches that of $SU(2)_L$ in the standard model. Further, the electromagnetic coupling obeys $e=g s_W$, where $s_W$ is the sine of the Weinberg angle. It follows that 
\be t_X = \fr{\sqrt{3}s_W}{\sqrt{3-4s^2_W}}.\ee Hence, the photon field can be rewritten as 
\be A=s_W A_3 +c_W \left(\fr{t_W}{\sqrt{3}}A_8+\sqrt{1-\fr{t^2_W}{3}}B\right).\ee The $Z$ boson is defined, orthogonal to $A$, i.e.
\be Z=c_W A_3 -s_W\left(\fr{t_W}{\sqrt{3}}A_8+\sqrt{1-\fr{t^2_W}{3}}B\right), \ee as in the standard model. A new $Z'$ boson is given, orthogonal to both $A$ and $Z$, i.e. to the hypercharge field in parentheses, as 
\be Z'=\sqrt{1-\fr{t^2_W}{3}}A_8 - \fr{t_W}{\sqrt{3}}B.\ee      

In the new basis $(A,Z,Z',C)$, the photon field is decoupled as a physical field. There only mix among $(Z,Z',C)$ given through a mass matrix in such basis as   
 \be  M'^2 =\begin{pmatrix}
m^2_Z & m^2_{ZZ'}& m^2_{ZC}\\
m^2_{ZZ'}& m^2_{Z'}& m^2_{Z'C}\\
m^2_{ZC}& m^2_{Z'C}& m^2_{C}\\ 
\end{pmatrix},\ee where 
\bea && m^2_{Z}= \fr{g^2}{4c^2_W}(u^2+v^2+4\kappa^2),\\
&& m^2_{ZZ'} =\fr{g^2}{4c^2_W\sqrt{3-4s^2_W}}\left[u^2-c_{2W}(v^2+4\kappa^2)\right],\\
&& m^2_{ZC}=\fr{g^2 t_N}{6c_W}(-u^2+v^2+4\kappa^2),\\
&& m^2_{Z'}=\fr{g^2}{4c^2_W(3-4s^2_W)}\left[u^2+c^2_{2W}(v^2+4\kappa^2)+4c^4_W(w^2+4\La^2)\right],\\
&& m^2_{Z'C}=-\fr{g^2t_N}{6c_W\sqrt{3-4s^2_W}}\left[u^2+c_{2W}(v^2+4\kappa^2)+4c^2_W (w^2-2\La^2) \right],\\
&& m^2_{C}=\fr{g^2 t^2_N}{9}(u^2+v^2+4\kappa^2+4w^2+4\La^2).
\eea 

Because of $u,v,\kappa\ll w,\La$, we have $m^2_Z,m^2_{ZZ'},m^2_{ZC}\ll m^2_{Z'},m^2_{Z'C}, m^2_C$. The matrix $M'^2$ is diagonalized via seesaw, yielding a physical light state $Z_1\simeq Z-\mathcal{E}_1Z'-\mathcal{E}_2 C$ with mass 
\be m^2_{Z_1}\simeq m^2_Z-\begin{pmatrix}\mathcal{E}_1 & \mathcal{E}_2\end{pmatrix}\begin{pmatrix}
m^2_{ZZ'}\\
m^2_{ZC}\end{pmatrix},  
\ee where the mixing parameters are
\be \begin{pmatrix}\mathcal{E}_1 & \mathcal{E}_2\end{pmatrix}
=\begin{pmatrix}
m^2_{ZZ'} & m^2_{ZC}\end{pmatrix}\begin{pmatrix}
m^2_{Z'}& m^2_{Z'C}\\
m^2_{Z'C}& m^2_{C} 
\end{pmatrix}^{-1}.\ee  
The relevant heavy states $\mathcal{Z'}\simeq \mathcal{E}_1 Z+Z'$ and $\mathcal{C}\simeq \mathcal{E}_2 Z +C$ possess a mass matrix,   
\be M''^2\simeq \begin{pmatrix}
m^2_{Z'}& m^2_{Z'C}\\
m^2_{Z'C}& m^2_{C} 
\end{pmatrix},\ee which yields two physical bosons, by diagonalization, such as 
\be Z_2= c_\theta \mathcal{Z}'-s_\theta\mathcal{C},\hs Z_3 = s_\theta \mathcal{Z}'+c_\theta\mathcal{C},\ee
where the mixing angle and their masses are given, respectively, by
 \be   t_{2\theta} = \fr{2m^2_{Z'C}}{m^2_C-m^2_{Z'}}\simeq \fr{12t_N \sqrt{3-t^2_W}(2\La^2-w^2)}{4t^2_N(3-t^2_W)(w^2+\La^2)-9(w^2+4\La^2)},\label{tta}\ee
 \bea 
  m^2_{Z_2,Z_3} &=&\fr 1 2 \left[m^2_{Z'}+m^2_C\mp \sqrt{(m^2_{Z'}-m^2_C)^2+4m^4_{Z'C}}\right]\crn
 &\simeq& \fr{g^2}{18(3-t^2_W)}\left\{9(w^2+4\La^2)+4t^2_N(3-t^2_W)(w^2+\La^2)\right.\crn
 &&\left.\mp \sqrt{[9(w^2+4\La^2)-4t^2_N(3-t^2_W)(w^2+\La^2)]^2+144t^2_N(3-t^2_W)(w^2-2\La^2)^2}\right\}.\label{mz2z3}\eea 
 
As we can be seen, because of $u,v,\kappa\ll w,\La$, the mixing angle $\theta$ and masses $m_{Z_2}, m_{Z_3}$ are governed by $w,\La$. For the case of interest $w\sim \La$, the new gauge bosons $\mathcal{Z}'$ and $\mathcal{C}$ finitely mix, i.e. $\theta \sim \mathcal{O}(1)$, and their physical states gain a mass at $w,\La$ scale, i.e. $m_{Z_2,Z_3}\sim w,\La$. Furthermore, $Z$ slightly mix with the new gauge bosons $Z'$ and $C$, due to 
\bea && \mathcal{E}_1\simeq \fr{\sqrt{3-4s^2_W}}{36c^4_W}\left[\fr{s^2_W(u^2+v^2+4\kappa^2)}{\La^2}+\fr{(3-2s^2_W)u^2-(3-4s^2_W)(v^2+4\kappa^2)}{w^2}\right]\ll 1,\\
&&\mathcal{E}_2\simeq \fr{1}{24t_N c^3_W}\left[\fr{s^2_W(u^2+v^2+4\kappa^2)}{\La^2}-2\fr{(3-2s^2_W)u^2-(3-4s^2_W)(v^2+4\kappa^2)}{w^2}\right]\ll 1,
\eea which are suppressed by $(u,v)^2/(w,\La)^2$, where $\kappa$ insignificantly contributes, since $\kappa\ll u,v$. Because $Z'\simeq \mathcal{Z}'-\mathcal{E}_1 Z$ and $C\simeq \mathcal{C}-\mathcal{E}_2 Z$ couple to fermions, they modify the well-measured couplings of $Z$ with fermions by amounts proportional to $\mathcal{E}_{1,2}$. The precision electroweak test gives a strong constraint on these mixing parameters, $\mathcal{E}_{1,2}\sim 10^{-3}$ \cite{pdg}. Taking $s^2_W\simeq 0.231$, $u^2+v^2\simeq (246\ \mathrm{GeV})^2$ (which is derived from the $W$ mass for $\kappa\ll u,v$), $t_N\sim 1$, and $w,\La\sim 4.5$ TeV, we estimate $\mathcal{E}_1\lesssim 10^{-3}$ and $\mathcal{E}_2\lesssim 10^{-3}$, which agree with the given limit. 

On the other hand, the mixing of $Z$ with $(Z',C)$ as well as the existence of a scalar triplet vacuum $\kappa\ll u,v$ modify the $\rho$-parameter, such as
\bea \Delta \rho &=& \fr{m^2_W}{c^2_W m^2_{Z_1}}-1\simeq -\fr{2\kappa^2}{u^2+v^2}+\mathcal{E}_1\fr{m^2_{ZZ'}}{m^2_Z}+\mathcal{E}_2\fr{m^2_{ZC}}{m^2_Z}\crn
&\simeq & -\fr{2\kappa^2}{u^2+v^2}+\fr{u^2-c_{2W}v^2}{36c^4_W (u^2
+v^2)}\left[\fr{s^2_W(u^2+v^2+4\kappa^2)}{\La^2}+\fr{(3-2s^2_W)u^2-(3-4s^2_W)(v^2+4\kappa^2)}{w^2}\right]\crn
&&+\fr{v^2-u^2}{36c^2_W(u^2+v^2)}\left[\fr{s^2_W(u^2+v^2+4\kappa^2)}{\La^2}-2\fr{(3-2s^2_W)u^2-(3-4s^2_W)(v^2+4\kappa^2)}{w^2}\right].\eea Because of $m^2_{ZZ'}\sim m^2_{ZC}\sim m^2_Z$, the $\rho$-parameter deviation is substantially suppressed by $\kappa^2/(u,v)^2\ll 1$ and $\mathcal{E}_{1,2}\sim (u,v)^2/(w,\La)^2\ll 1$. With the values of parameters as given, i.e. $s^2_W\simeq 0.231$, $u^2+v^2\simeq (246\ \mathrm{GeV})^2$, $t_N\sim 1$, and $w,\La\sim 4.5$, the $\rho$-parameter deviation agrees with the global fit $\Delta\rho\sim 3.1\times 10^{-4}$ \cite{pdg}, given that $\kappa\sim \mathcal{O}(1)$ GeV and an appropriate value of $u$ or $v$ (see below in detail when combining both $\Delta\rho$ and $\mathcal{E}_{1,2}$ for numerical investigation). It is noted that the $\rho$-parameter deviation above comes from the tree-level contribution by the nontrivial vacuum of and the mixing with the new particles. At the loop level, the $\rho$-parameter may receive radiative contributions to the relevant self-energy diagrams by non-degenerate doublets of new particles, such as the gauge boson doublet $(X,Y)$ and the lepton doublets $(E^0,E^-)$. However, these corrections are strongly suppressed by $(u,v,\kappa,\mu)^2/(w,\La)^2$ and loop factors $1/16\pi^2$, which are not significant than the tree-level one and possibly skipped in this work \cite{331rho1,331rho2,331rho3,d1e4}.    

\begin{figure}[h]
\bc
\includegraphics[scale=1]{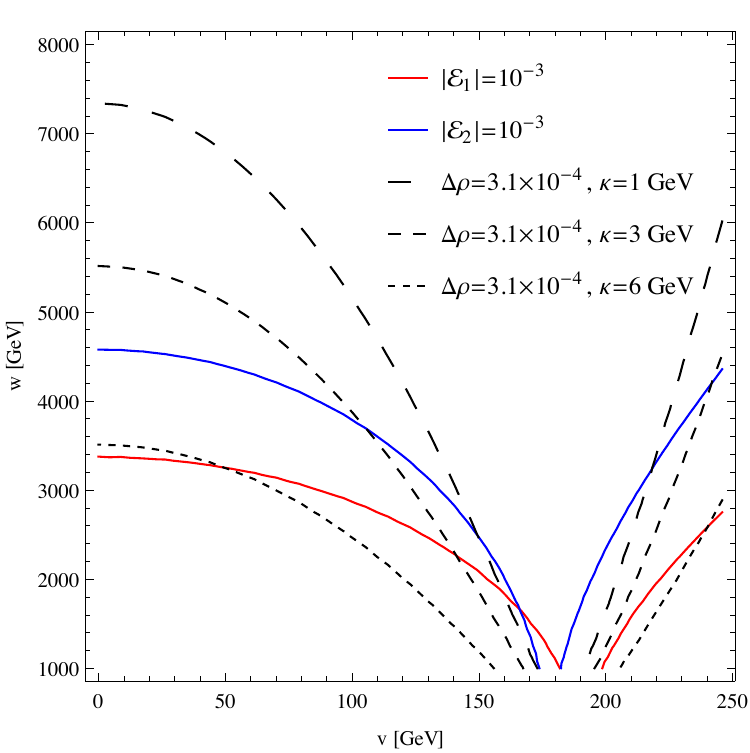}
\caption[]{\label{fig1} The mixing parameters $\mathcal{E}_{1,2}$ and the $\rho$-parameter deviation $\Delta \rho$ contoured according to the current limits as functions of $v,w$ for $\kappa=1$, 3, and 6 GeV.}
\ec
\end{figure}
Explicitly, the viable parameter regime $(v,w)$ can be obtained by contouring the mixing parameters according to the precision electroweak test, say $\mathcal{E}_{1,2}=10^{-3}$, and the $\rho$-parameter deviation according to the global fit, i.e. $\Delta\rho=3.1\times 10^{-4}$, as depicted in Fig. \ref{fig1}. Note that $u$ is related to $v$ as $u=\sqrt{(246\ \mathrm{GeV})^2-v^2}$. We take $t_N=\sqrt{3}/2$ which results from $N=(2/\sqrt{3})T_{8R}$, thus $1/g_N^2=4/3g_R^2$, by normalizing the relevant fields similar to the case of photon, as well as putting $g_R=g$.\footnote{They would be different due to a running from the trinification to the 3-3-1-1 by RGE. However, the difference is almost insignificant for this study and is omitted.} It is clear from the figure that $\Delta\rho$ is sensitive to $\kappa$, while $\mathcal{E}_{1,2}$ are not. Additionally, we have taken $\La=w$. In fact, in the $\mathcal{E}_{1,2}$ and $\Delta\rho$ expressions, the terms associated with $\La$ proportional to $s^2_W$ are more suppressed than the terms associated with $w$, given that $\La\sim w$. Hence, $\mathcal{E}_{1,2}$ and $\Delta\rho$ depend dominantly on the terms of $w$. The contours as given slightly change when $\La$ changes but remains the condition $\La\sim w$. Hence, the figure can apply for $\La\sim w$, up to insignificant small modifications. Since $\mathcal{E}_{1,2}$ and $\Delta\rho$ are governed by the terms of $w$ (i.e., the second term in the square brackets), the effect of closing 3-3-1-1 model at the weak scale occurs. Indeed, when the nominator of the $w$ terms tends to zero, i.e. the mixing of $Z$ and $(Z',C)$ vanishes, the new physics scale $w$ is reduced below 1 TeV close to the weak scale, as seen from the figure.\footnote{Although the standard model is not modified by closing the 3-3-1-1 symmetry at the weak scale, as the electroweak measurements bearing from gauge boson mixing and $\rho$-parameter are ensured, the high energy collider definitely limits the new particle masses, requiring that $(w,\La)$ must be beyond 1 TeV.} That point is set by \be (3-2s^2_W)u^2=(3-4s^2_W)v^2,\ee leading to $v=182.5$ GeV, as predicted in the figure. This phenomenon is also observed in the 3-3-1 model \cite{close331,close331p}. Lastly, the available parameter regime of $w$ can be achieved so that it lies above all the given curves.      

\subsection{Scalar mass}

The scalar potential is given by
\bea V&=&\mu^2_1 \rho^\dagger \rho +\mu^2_2 \varphi^\dagger \varphi +\mu^2_3 \eta^\dagger \eta \crn
&&+ \la_1 (\rho^\dagger \rho)^2+\la_2 (\varphi^\dagger \varphi)^2+\la_3 (\eta^\dagger \eta)^2\crn
&&+\la_4 (\rho^\dagger \rho)(\varphi^\dagger \varphi)+\la_5 (\rho^\dagger \rho)(\eta^\dagger \eta) +\la_6 (\varphi^\dagger \varphi)(\eta^\dagger \eta)\crn
&&+\la_7 (\rho^\dagger \varphi)(\varphi^\dagger \rho)+\la_8 (\rho^\dagger \eta)(\eta^\dagger \rho)+\la_9 (\varphi^\dagger \eta)(\eta^\dagger \varphi)\crn
&&+\mu^2_4\mathrm{Tr}(\sigma^\dagger \sigma) + \la_{10} \mathrm{Tr}^2(\sigma^\dagger \sigma) +\la_{11} \mathrm{Tr} (\sigma^\dagger \sigma)^2\crn
&&+ (\la_{12}\rho^\dagger \rho +\la_{13}\varphi^\dagger \varphi+\la_{14}\eta^\dagger \eta)\mathrm{Tr}(\sigma^\dagger \sigma)\crn
&&+ \la_{15} (\rho^\dagger \sigma)(\sigma^\dagger \rho)+\la_{16} (\varphi^\dagger \sigma)(\sigma^\dagger \varphi)+ \la_{17} (\eta^\dagger \sigma)(\sigma^\dagger \eta)\crn
&& +(f_1\eta \rho \varphi+f_2 \rho^T\sigma^\dagger \rho+H.c.),\eea where the parameters $\mu_{1,2,3,4}$ and $f_{1,2}$ have a mass dimension, while the couplings $\la$'s are dimensionless. Let us expand the scalar fields around their VEVs, such as
\bea &&\eta = \begin{pmatrix} \fr{u}{\sqrt{2}}+\fr{R_1+i I_1}{\sqrt{2}}\\
\eta^-_2\\
\eta^-_3
\end{pmatrix},\hs \rho = \begin{pmatrix} \rho^+_1\\
\fr{v}{\sqrt{2}} + \fr{R_2+i I_2}{\sqrt{2}}\\
\fr{R'_3+iI'_3}{\sqrt{2}}
\end{pmatrix},\\ 
&&\varphi = \begin{pmatrix} \varphi^+_1\\
\fr{R'_2+iI'_2}{\sqrt{2}}\\
\fr{w}{\sqrt{2}}+\fr{R_3+i I_3}{\sqrt{2}}
\end{pmatrix},\hs  \sigma = \begin{pmatrix}\sigma^{++}_{11} & \fr{\sigma^{+}_{12}}{\sqrt{2}} & \fr{\sigma^{+}_{13}}{\sqrt{2}} \\
\fr{\sigma^{+}_{12}}{\sqrt{2}}  & \fr{\kappa}{\sqrt{2}}+\fr{R_4+iI_4}{\sqrt{2}} & \fr{R'_5+iI'_5}{2}  \\
\fr{\sigma^{+}_{13}}{\sqrt{2}}  & \fr{R'_5+i I'_5}{2} & \fr{\La}{\sqrt{2}}+\fr{R_6+iI_6}{\sqrt{2}}
\end{pmatrix}.\eea

Substituting the scalar field expansions into the potential, the gauge invariance requires the linear terms in fields vanished, yielding minimization conditions,
\bea
&& u \left[ \lambda_{14} (\kappa^2 + \Lambda^2) + 2 \mu_3^2 + 2 \lambda_3 u^2 + \lambda_6 w^2 \right]+\lambda_5 u v^2 - \sqrt{2} f_1 v w  = 0 ,\crn
&& v \left[ \kappa^2 (\lambda_{12} + \lambda_{15}) + \lambda_{12} \Lambda^2 + 2 \mu_1^2 + \lambda_5 u^2 + \lambda_4 w^2 \right]+2\sqrt{2} f_2 \kappa v - \sqrt{2} f_1 u w + 2\lambda_1 v^3 = 0, \crn
&&w \left[ \kappa^2 \lambda_{13} + (\lambda_{13} + \lambda_{16}) \Lambda^2 + 2 \mu_2^2 + \lambda_6 u^2 + \lambda_4 v^2 + 2 \lambda_2 w^2  \right] -\sqrt{2} f_1  u v= 0, \crn
&& 2 (\lambda_{10} + \lambda_{11}) \Lambda^2 + 2 \mu_4^2 + \lambda_{14} u^2 + \lambda_{12} v^2 + (\lambda_{13} + \lambda_{16}) w^2+2\lambda_{10} \kappa^2   = 0,\crn
&&  \kappa \left[2 \lambda_{10} \Lambda^2 + 2 \mu_4^2 + \lambda_{14} u^2 + (\lambda_{12} + \lambda_{15}) v^2 + \lambda_{13} w^2 \right] + \sqrt{2} f_2 v^2 +2\kappa^3 (\lambda_{10} + \lambda_{11}) = 0.\nn
\eea These constraints give a nontrivial solution for $(\kappa,u,v,w,\La)$, provided that $\mu^2_{1,2,3,4}<0$ and the conditions for scalar couplings, e.g. $\la_{1,2,3},\la_{10}+\la_{11}>0$, so that the quartic coupling matrix is copositive, responsible for vacuum stability \cite{vacuumstability}. The hierarchy $u,v\ll w,\La$ acquires $|\mu_1|, |\mu_3|\ll |\mu_2|,|\mu_4|$. Whereas, the hierarchy $\kappa\ll u,v$ is manifestly a result of the last minimization condition, i.e. \be \kappa\simeq -\fr{\sqrt{2}f_2v^2}{2\la_{10}\La^2+2\mu^2_4+\la_{13}w^2}\simeq \fr{\sqrt{2}f_2v^2}{2\la_{11}\La^2+\la_{16}w^2}\sim \fr{f_2}{(w,\La)}\fr{v^2}{(w,\La)},\ee which is suppressed by a seesaw, even for the soft term $f_2\sim w,\La$.

The quadratic terms in fields in the scalar potential provide physical scalar mass spectrum. The CP-even neutral scalars $(R_1,R_2,R_3,R_4,R_6)$ obtain a mass matrix in such basis as
\bea
M^2_{R}=
\begin{pmatrix}
	2 \lambda_3 u^2 + \frac{f_1 v w}{\sqrt{2} u} & \lambda_5 u v - \frac{f_1 w}{\sqrt{2}} & -\frac{f_1 v}{\sqrt{2}} + \lambda_6 u w & \lambda_{14}\kappa  u & \lambda_{14} \Lambda u ,\\
	\lambda_5 u v - \frac{f_1 w}{\sqrt{2}} & 2 \lambda_1 v^2 + \frac{f_1 u w}{\sqrt{2} v} & -\frac{f_1 u}{\sqrt{2}} + \lambda_4 v w & \mathcal{A} & \lambda_{12} \Lambda v, \\
	-\frac{f_1 v}{\sqrt{2}} + \lambda_6 u w & -\frac{f_1 u}{\sqrt{2}} + \lambda_4 v w & \frac{f_1 u v}{\sqrt{2} w} + 2 \lambda_2 w^2 & \lambda_{13} \kappa w & (\lambda_{13} + \lambda_{16}) \Lambda w \\
 \lambda_{14} 	\kappa u & \mathcal{A} &  \lambda_{13}\kappa w & \mathcal{B} & 2  \lambda_{10} \kappa \Lambda \\
	\lambda_{14} \Lambda u & \lambda_{12} \Lambda v & (\lambda_{13} + \lambda_{16}) \Lambda w & 2  \lambda_{10} \kappa \Lambda & 2 (\lambda_{10} + \lambda_{11}) \Lambda^2
\end{pmatrix},
\eea
where
\bea
	\mathcal{A} &\equiv& \frac{\kappa}{v} \left( -2 \kappa^2 \lambda_{11} + 2 \lambda_{11} \Lambda^2 + \lambda_{12} v^2 + \lambda_{16} w^2 \right), \\
	\mathcal{B} &\equiv& \frac{1}{2} \left[\kappa^2 (4 \lambda_{10} + 6 \lambda_{11}) - 2 \lambda_{11} \Lambda^2 + \lambda_{15} v^2 - \lambda_{16} w^2 \right].
\eea
We assume $f_1\sim (w,\La)$ for this kind of the model, as usual. At the effective limit, $\kappa \ll u, v \ll f_1, w, \Lambda$, the field groups $(R_1,R_2)$, $(R_3,R_6)$, and $R_4$ are in each pair of such groups decoupled (do not mix). However, there are mixings within each group, which contains more than one field. That said, within this VEV regime, the physical eigenstates can be approximated, related to the gauge states, such as
\be
	H \simeq c_\beta R_1 + s_\beta R_2,\hs
	H_1 \simeq -s_\beta R_1 + c_\beta R_2,
\ee
\be
	H_2 \simeq c_\xi R_3 + s_\xi R_6,\hs
	H_3 \simeq -s_\xi  R_3 + c_\xi  R_6, 
	\ee and $H_4 \simeq  R_4$, where the mixing angles are defined by 
\be t_\beta =\fr{v}{u},\hs t_{2\xi}=\fr{(\lambda_{13} + \lambda_{16}) w \Lambda}{ [\lambda_2 w^2-(\lambda_{10} + \lambda_{11}) \Lambda^2]}.\ee The corresponding physical masses are given by
\bea
m_{H}^2 & \simeq  & \frac{2 \left( \lambda_3 u^4 + \lambda_5 u^2 v^2 + \lambda_1 v^4 \right)}{u^2 + v^2}, \hs m_{H_1}^2 \simeq \frac{f_1 (u^2 + v^2) w}{\sqrt{2} u v},  \crn
m_{H_2}^2 & \simeq& (\lambda_{10} + \lambda_{11}) \Lambda^2 + \lambda_2 w^2 
\crn
&& - \sqrt{(\lambda_{10} + \lambda_{11})^2 \Lambda^4 + \left[(\lambda_{13} + \lambda_{16})^2 - 
	2 (\lambda_{10} + \lambda_{11}) \lambda_2 \right] \Lambda^2 w^2 + \lambda_2^2 w^4}, \crn 
	m_{H_3}^2 & \simeq& (\lambda_{10} + \lambda_{11}) \Lambda^2 + \lambda_2 w^2 \crn
	&&+
	\sqrt{(\lambda_{10} + \lambda_{11})^2 \Lambda^4 + \left[(\lambda_{13} + \lambda_{16})^2 - 
	2 (\lambda_{10} + \lambda_{11}) \lambda_2 \right] \Lambda^2 w^2 + \lambda_2^2 w^4}, \nonumber  \\
	m_{H_4}^2 &\simeq& \frac{1}{2} \left( -2 \lambda_{11} \Lambda^2 + \lambda_{15} v^2 - \lambda_{16} w^2 \right) \nonumber .
\eea It is clear that the standard model Higgs boson like $H$ gains a mass at the weak scale $m_H\sim (u,v)$. By contrast, all other CP-even neutral Higgs bosons $H_{1,2,3,4}$ are new fields with respective masses at $(w,\La)$ scale. 

The CP-odd neutral scalar sector includes $(I_1, I_2, I_3, I_4, I_6)$. Among them, the last field, $I_6$, is decoupled from the remainders and is massless, hence it identified as the Goldstone boson absorbed by $Z_3$, i.e. $G_{Z_3}\equiv I_6$. The remaining fields obtain a mass matrix given in the basis $(I_1, I_2, I_3,I_4)$ as
\be
	M^2_{I} = \begin{pmatrix}
		\frac{f_1 v w}{\sqrt{2} u} & \frac{f_1 w}{\sqrt{2}} & \frac{f_1 v}{\sqrt{2}} & 0 \\
		\frac{f_1 w}{\sqrt{2}} & \frac{ \sqrt{2} f_1 u v w - 4 \kappa^2 \mathcal{D}}{2 v^2} & \frac{f_1 u}{\sqrt{2}} & \frac{\kappa}{v} D \\
		\frac{f_1 v}{\sqrt{2}} & \frac{f_1 u}{\sqrt{2}} & \frac{f_1 u v}{\sqrt{2} w} & 0 \\
		0 & \frac{\kappa}{v} D & 0 & -\frac{1}{2} D
	\end{pmatrix}. \label{eq:CPoddMatrix}
\ee
where $D \equiv 2 \lambda_{11}(\Lambda^2-\kappa^2) - \lambda_{15} v^2 + \lambda_{16} w^2$. 	
This matrix has two zero eigenvalues with corresponding eigenstates, called $\mathcal{A}_{1,2}$, identified as the Goldstone bosons coupled to $Z_{1,2}$, i.e. $G_{Z_{1,2}}\equiv \mathcal{A}_{1,2}$, respectively. The remaining pseudoscalars, called $\mathcal{A}_{3,4}$, are massive.  The physical eigenstates are related to the gauge states as
\bea
	\mathcal{A}_1 &\simeq& c_\beta I_1 - s_\beta I_2, \\
	 \mathcal{A}_2 &\simeq& -s_\zeta \left(s_\beta I_1+c_\beta I_2 \right) +c_\zeta I_3, \\
	\mathcal{A}_3&\simeq& c_\zeta \left(s_\beta I_1+c_\beta I_2 \right) +s_\zeta I_3, \\
	 \mathcal{A}_4 &\simeq& I_4, 
\eea
where $t_\zeta =\frac{uv}{w\sqrt{u^2+v^2}}$, with respective masses, $m_{\mathcal{A}_1}=0$, $m_{\mathcal{A}_2}=0$, \bea  && m_{\mathcal{A}_3}^2 =\frac{f_1}{\sqrt{2}} \left(\fr{v w}{u} + \fr{u v}{w} + \fr{uw}{v}\right), \\ 
&& m_{\mathcal{A}_4}^2= -\frac{1}{2} \left( 2 \lambda_{11} \Lambda^2 - \lambda_{15} v^2 + \lambda_{16} w^2 \right), \eea Except for the massless Goldstone bosons $G_{Z_{1,2,3}}$, the pseudoscalars $\mathcal{A}_{3,4}$ are physical fields with masses at $(w,\La)$ scale. 

Concerning charged scalars, the single doubly-charged scalar $\sigma^{\pm\pm}_{11}$ is a physical field by itself with a mass, given by \be m^2_{\sigma_{11}}=\fr 1 2 (-2\la_{11}\La^2+\la_{17} u^2-\la_{16} w^2),\ee which is at $(w,\La)$ scale. By contrast, the singly-charged and (matter) parity-even scalars, say $(\eta_{2}^\pm, \rho_{1}^\pm, \sigma_{12}^\pm)$, mix via a mass matrix given in such basis as
\be
M^2_{\mathrm{c.e.}}=\begin{pmatrix}
	D_{11} & \frac{1}{2} (\lambda_8 u v + \sqrt{2} f_1 w) & \frac{\lambda_{17}\kappa  u}{2 \sqrt{2}} \\
	\frac{1}{2} (\lambda_8 u v + \sqrt{2} f_1 w) & D_{22} & \frac{\kappa ( 4 \lambda_{11} \Lambda^2-4  \lambda_{11} \kappa^2 - \lambda_{15} v^2 + 2 \lambda_{16} w^2)}{2 \sqrt{2} v} \\
	\frac{\lambda_{17}\kappa  u}{2 \sqrt{2}} & \frac{\kappa (4 \lambda_{11} \Lambda^2 -4  \lambda_{11}\kappa^2 - \lambda_{15} v^2 + 2 \lambda_{16} w^2)}{2 \sqrt{2} v} & D_{33}
\end{pmatrix},\nn
\ee
where 
\bea
	D_{11} &\equiv& \frac{\kappa^2 \lambda_{17} u + \lambda_8 u v^2 + \sqrt{2} f_1 v w}{2 u} , \nonumber \\
	D_{22} &\equiv&\frac{4 \kappa^4 \lambda_{11} + u v (\lambda_8 u v + \sqrt{2} f_1 w) - \kappa^2 (4 \lambda_{11} \Lambda^2 - \lambda_{15} v^2 + 2 \lambda_{16} w^2)}{2 v^2}, \nonumber \\
	D_{33} &\equiv& \frac{1}{4} \left[4\lambda_{11} ( \kappa^2 - \Lambda^2 )+ \lambda_{17} u^2 + \lambda_{15} v^2 - 2 \lambda_{16} w^2\right]. \nonumber
\eea Diagonalizing this mass matrix, we identify a massless eigenstate to be the Goldstone boson of $W^\pm$, such as
\bea
G^\pm_W &=& \frac{u}{\sqrt{2 \kappa^2 + u^2 + v^2}} \eta_{2}^\pm
-\frac{v}{\sqrt{2 \kappa^2 + u^2 + v^2}} \rho_{1}^\pm  
-\frac{\sqrt{2} \kappa}{\sqrt{2 \kappa^2 + u^2 + v^2}} \sigma_{12}^\pm,\crn
&\simeq& -s_\beta \rho^\pm_1 + c_\beta \eta^\pm_2.
\eea The remaining eigenstates determine two physical massive singly-charged Higgs, say
\bea
H^\pm_1 &=& \frac{u v}{D_1} \eta_{2}^\pm + \frac{2 \kappa^2 + u^2}{D_1} \rho_{1}^\pm - \frac{\sqrt{2} \kappa v}{D_1} \sigma_{12}^\pm\simeq c_\beta \rho^\pm_1+s_\beta \eta^\pm_2, \\
H_2^\pm &=& \frac{\sqrt{2} \kappa}{\sqrt{2 \kappa^2 + u^2}} \eta_{2}^\pm + \frac{u}{\sqrt{2 \kappa^2 + u^2}} \sigma_{12}^\pm \simeq \sigma^\pm_{12},
\eea
where $D_1 \equiv \sqrt{(2 \kappa^2 + u^2)(2 \kappa^2 + u^2 + v^2)} $, with respective masses,
\bea
m^2_{H^\pm_1} &\simeq & \frac{(u^2 + v^2) (\lambda_8 u v + \sqrt{2} f_1 w)}{2 u v}, \\
m^2_{H_2^\pm} &\simeq & \frac{1}{4} \left(-4 \lambda_{11} \Lambda^2 + \lambda_{17} u^2 + \lambda_{15} v^2 - 2 \lambda_{16} w^2\right),
\eea which are at $(w,\La)$ scale. 

The singly-charged and (matter) parity-odd scalars, say $(\eta_3^\pm, \varphi_1^\pm, \sigma_{13}^\pm)$, mix via a mass matrix given in such basis as  
\bea
M^2_{\mathrm{c.o.}}=\begin{pmatrix}
	\frac{\lambda_{17} \Lambda^2 u + \sqrt{2} f_1 v w + \lambda_9 u w^2}{2 u} & \frac{1}{2} (\sqrt{2} f_1 v + \lambda_9 u w) & \frac{\lambda_{17} \Lambda u}{2 \sqrt{2}} \\
	\frac{1}{2} (\sqrt{2} f_1 v + \lambda_9 u w) & \frac{\sqrt{2} f_1 u v - \lambda_{16} \Lambda^2 w + \lambda_9 u^2 w}{2 w} & \frac{\lambda_{16} \Lambda w}{2 \sqrt{2}} \\
	\frac{\lambda_{17} \Lambda u}{2 \sqrt{2}} & \frac{\lambda_{16} \Lambda w}{2 \sqrt{2}} & \frac{1}{4} (\lambda_{17} u^2 - \lambda_{16} w^2)
\end{pmatrix}.
\eea
This mass matrix has a massless eigenstate, identified as the Goldstone boson of the $X^\pm$ gauge boson, such as
\bea
	G^\pm_X &=& -\frac{u}{\sqrt{2 \Lambda^2 + u^2 + w^2}} \eta_{3}^\pm + \frac{w}{\sqrt{2 \Lambda^2 + u^2 + w^2}} \varphi_1^\pm + \frac{\sqrt{2} \Lambda}{\sqrt{2 \Lambda^2 + u^2 + w^2}} \sigma_{13}^\pm,\crn
	&\simeq& s_\al \varphi^\pm_1+c_\al \sigma^\pm_{13},
\eea where $t_\al=w/\sqrt{2}\La$.
The remaining eigenstates define two physical massive singly-charged parity-odd Higgs,
\bea
	H'^{\pm}_1 &=& c_\delta \left( \frac{\sqrt{2} \Lambda}{D_2} \eta_{3}^\pm + \frac{u}{D_2} \sigma_{13}^\pm \right) 
	- s_\delta \left( \frac{u w}{D_3} \eta_{3}^\pm + \frac{2 \Lambda^2 + u^2}{D_3} \varphi_1^\pm - \frac{\sqrt{2} \Lambda w}{D_3} \sigma_{13}^\pm \right)\crn
	&\simeq& \eta^\pm_3, \\
	H'^{\pm}_2 &=& s_\delta \left( \frac{\sqrt{2} \Lambda}{D_2} \eta_{3}^\pm + \frac{u}{D_2} \sigma_{13}^\pm \right) 
	+ c_\delta \left( \frac{u w}{D_3} \eta_{3}^\pm + \frac{2 \Lambda^2 + u^2}{D_3} \varphi_1^\pm - \frac{\sqrt{2} \Lambda w}{D_3} \sigma_{13}^\pm \right)\crn
	&\simeq& c_\al \varphi^\pm_1 -s_\al \sigma^\pm_{13},
\eea
where $D_2 \equiv \sqrt{2 \Lambda^2 + u^2}$, 
$D_3 \equiv \sqrt{(2 \Lambda^2 + u^2)(2 \Lambda^2 + u^2 + w^2)}$,
and the mixing angle is 
\bea
s_{2\delta} &=& \frac{2 u w\La (2 \Lambda^2 + u^2)^{-3/2} [4 f_1 v + \sqrt{2} (\lambda_{16} + 2 \lambda_9) u w] \sqrt{(2 \Lambda^2 + u^2) (2 \Lambda^2 + u^2 + w^2)}}{2 \sqrt{2} f_1 v (u^2 + w^2) + u w [\lambda_{17} (2 \Lambda^2 + u^2) - \lambda_{16} (2 \Lambda^2 + w^2) + 2 \lambda_9 (u^2 + w^2)]}\crn
&\sim&(u,v)/(w,\La)\ll 1.
\eea The corresponding masses of the Higgs $H'^{\pm}_1$ and $H'^{\pm}_2$ are approximated by
\bea
m^2_{H'^{\pm}_1} &\simeq& \frac{\lambda_{17} \Lambda^2 u + \sqrt{2} f_1 v w + \lambda_9 u w^2}{2 u},  \\
m^2_{H'^{\pm}_2} &\simeq& \frac{\left(\sqrt{2} f_1 u v - \lambda_{16} \Lambda^2 w + \lambda_9 u^2 w\right) (2 \Lambda^2 + w^2)}{4 \Lambda^2 w},
\eea which are at $(w,\La)$ scale.

The CP-even electrically-neutral and (matter) parity-odd scalar fields $(R'_5, R'_2, R'_3)$ mix according to the mass matrix given in such basis as
\bea
M^{2}_{R'}=\begin{pmatrix}
	\kappa \lambda_{11} (\kappa + \Lambda) + \frac{1}{4} (\lambda_{15} v^2 - \lambda_{16} w^2) & \frac{\lambda_{16} (\kappa + \Lambda) w}{2 \sqrt{2}} & E_{13} \\
	\frac{\lambda_{16} (\kappa + \Lambda) w}{2 \sqrt{2}} & \frac{\sqrt{2} f_1 u v + \lambda_{16} (\kappa^2 - \Lambda^2)  w + \lambda_7 v^2 w}{2 w} & \frac{1}{2} (\sqrt{2} f_1 u + \lambda_7 v w) \\
E_{13} & \frac{1}{2} (\sqrt{2} f_1 u + \lambda_7 v w) & E_{33}
\end{pmatrix},\nn
\eea
where
\bea
E_{13}&\equiv &\frac{4 \lambda_{11} \kappa ( \Lambda^2-\kappa^2)  - \lambda_{15} v^2(\kappa+\Lambda)+ 2 \kappa \lambda_{16} w^2}{2 \sqrt{2} v}, \nonumber \\
E_{33}&\equiv& \frac{1}{2} \left\{\lambda_{15} (\kappa - \Lambda)^2 + \frac{\sqrt{2} f_1 u w}{v} + \lambda_7 w^2 + \frac{2 \kappa (\kappa - \Lambda) [2 \lambda_{11} (\kappa ^2- \Lambda^2)  - \lambda_{16} w^2]}{v^2}\right\} \nonumber .
\eea
Diagonalizing this mass matrix reveals one massless eigenstate and two massive eigenstates. The massless eigenstate is given by
\bea
S'_2 &=& \frac{\sqrt{2} \Lambda}{\sqrt{2 \Lambda^2 + v^2 + w^2}} R'_5 +\frac{w}{\sqrt{2 \Lambda^2 + v^2 
		+ w^2}}R'_2 - \frac{v}{\sqrt{2 \Lambda^2 + v^2 + w^2}}R'_3\crn
		&\simeq& s_\al R'_2+c_\al R'_5.
\eea Defining intermediate states orthogonal to $S'_2$, such as 
\bea
&& S'_3  \equiv \frac{v}{\sqrt{v^2 + w^2}} R_2^\prime+ \frac{w}{\sqrt{v^2 + w^2}} R_3^\prime,\\
&& S'_5 \equiv \frac{-(v^2 + w^2)R_5^\prime+ \sqrt{2} \Lambda w R_2^\prime  -\sqrt{2} \Lambda v R_3^\prime}{\sqrt{(v^2+w^2)(v^2 + w^2+2 \Lambda^2 )}}, \eea
the two massive eigenstates are obtained by
\bea
&& H'_1  = c_\ga S'_3 - s_\ga S'_5 \simeq R'_3, \\ 
&& H'_2  = s_\ga S'_3 +c_\ga S'_5\simeq c_\al R'_2-s_\al R'_5, 
\eea
where the $S'_3$-$S'_5$ mixing angle is approximated as \bea 
t_{2\ga} &\simeq& \frac{4 (\lambda_{15} + \lambda_{16}) v^2 (\Lambda w)^{3/2}}{2 w^3 (\sqrt{2} f_1 u + \lambda_7 v w) - 2\lambda_{15}vw\La ( v^2 -  w\Lambda ) + \lambda_{16}v (v^2 w^2 -2v^2 \Lambda^2  + 2w^3 \Lambda )}\crn
&\sim& v/(w,\La)\ll 1. 
\eea
The corresponding masses are
\bea
m^2_{H_1^\prime } &\simeq & \frac{f_1 u (v^2 + w^2)}{\sqrt{2} v w} + \frac{ \lambda_{15} \Lambda^2 w^2 + \lambda_7 (v^2 + w^2)^2-\lambda_{16} \Lambda^2 v^2 }{2 (v^2 + w^2)},\\
m^2_{H_2^\prime }&\simeq & \frac{w^2 (v^2 + 2 \Lambda w) (\lambda_{15} v^2 - \lambda_{16} w^2)}{4 (v^2 + w^2)^2},
\eea which are all at $(w,\La)$ scale. 

The CP-odd electrically-neutral and (matter) parity-odd scalar sector includes $(I_5^\prime, I_2^\prime, I_3^\prime)$, mixing through the mass matrix constructed in such basis as 
\bea
M^2_{I'}=\begin{pmatrix}
	\frac{4 \lambda_{11}( \kappa^2 -  \kappa \Lambda )+ \lambda_{15} v^2 - \lambda_{16} w^2}{4} & -\frac{\lambda_{16} (\kappa - \Lambda) w}{2 \sqrt{2}} & (M^2_{I'})_{13} \\
	-\frac{\lambda_{16} (\kappa - \Lambda) w}{2 \sqrt{2}} & \frac{\sqrt{2} f_1 u v + \lambda_{16} w(\kappa^2  - \Lambda^2 ) + \lambda_7 v^2 w}{2 w} & -\frac{f_1 u}{\sqrt{2}} - \frac{\lambda_7 v w}{2} \\
(M^2_{I'})_{13} & -\frac{f_1 u}{\sqrt{2}} - \frac{\lambda_7 v w}{2} & (M^2_{I'})_{33}
\end{pmatrix} \nonumber
\eea
where
\bea
(M^2_{I'})_{13} &=&-\frac{4  \lambda_{11}\kappa (\kappa^2-\Lambda^2) + \lambda_{15}  v^2 (\kappa+\Lambda)- 2 \lambda_{16}\kappa w^2}{2 \sqrt{2} v} , \nonumber  \\
(M^2_{I'})_{33}&=& \fr{\lambda_{15} (\kappa + \Lambda)^2+ \lambda_7 w^2}{2} + \frac{ f_1 u w}{\sqrt{2}v} + \frac{ \kappa (\kappa + \Lambda) [2 \lambda_{11} (\kappa^2 - \Lambda^2) - \lambda_{16} w^2]}{v^2}. \nonumber
\eea
Similar to the real/scalar part, the imaginary/pseudoscalar part behaves the same. Indeed, diagonalizing this matrix, we obtain one massless eigenstate, identified as
\bea
P'_2 & =&\frac{\sqrt{2} \Lambda}{\sqrt{2 \Lambda^2 + v^2 + w^2}} I_5^\prime +\frac{w}{\sqrt{2 \Lambda^2 + v^2 
		+ w^2}}I_2^\prime - \frac{v}{\sqrt{2 \Lambda^2 + v^2 + w^2}}I_3^\prime,\crn 
&\simeq&s_\al I'_2+ c_\al I'_5.  \eea
Similar to the real/scalar part, defining intermediate states orthogonal to $P'_2$, such as 
\bea
 P^\prime_3 &\equiv & -\frac{v}{\sqrt{v^2 + w^2}} I'_2 + \frac{w}{\sqrt{v^2 + w^2}} I'_3, \nonumber  \\
 P^\prime_5 &\equiv &\frac{-(v^2 + w^2)I'_5 + \sqrt{2} \Lambda w I'_2+ \sqrt{2}  \Lambda v I'_3}{\sqrt{(v^2+w^2)(v^2 + w^2+2 \Lambda^2)}},  
\eea we obtain the remaining two massive eigenstates,
\bea
A_1^\prime & = &c_\ga P^\prime_3 -s_\ga P^\prime_5 \simeq I'_3, \\ 
A_2^\prime &=& s_\ga P_3^\prime +c_\ga P_5^\prime\simeq c_\al I'_2 - s_\al I'_5,
\eea where the mixing in imaginary part ($P'_3$-$P'_5$) is the same real part ($S'_3$-$S'_5$), determined by $\ga$. 
The masses of the physical pseudoscalars $A_1^\prime$ and $A_2^\prime$ are found to be
\bea
m^2_{A_1^\prime} \simeq m^2_{H_1^\prime}+ \mathcal{O}_1(f_2, w,\La)^2, \hspace{0.5 cm}
m^2_{A_2^\prime} \simeq m^2_{H_2^\prime}+ \mathcal{O}_2(\kappa, u,v)^2,
\eea for which the corresponding mass splittings are approximately given as in the previous section of radiative neutrino mass. In other words, the complex field $H'^0_1\equiv (H'_1+iA'_1)/\sqrt{2}$ is largely separated in mass with an amount at the $(w,\La)$ scale, while their masses are given also at $(w,\La)$ scale. By contrast, $H'^0_2=(H'_2+iA'_2)/\sqrt{2}$ is separated in mass proportional to the weak scale, while their masses at $(w,\La)$ scale. Because $H'^0_2$ is the neutral component of a weak scalar doublet (the remaining component is $H'^\pm_2$), their mass splitting is crucial for their real or imaginary part becoming a dark matter candidate evading the dark matter direct detection. Note that $H'^0_1$ does not interact with $Z$, thus the mass splitting or even degeneracy is not important. However, for small mass splitting (but unwanted parameter adjustment is needed), the dark matter coannihilation that sets the relic density is interesting.      

Last, but not least, the massless eigenstates $S'_2$ and $P'_2$ can be combined to form the Goldstone boson associated with $Y^0$ gauge boson, such as
\bea G^0_Y &=& \fr{S'_2 + i P'_2}{\sqrt{2}} \crn
&=& \frac{\sqrt{2} \Lambda}{\sqrt{2 \Lambda^2 + v^2 + w^2}} \sigma^0_{23} +\frac{w}{\sqrt{2 \Lambda^2 + v^2 
		+ w^2}}\varphi^0_2 - \frac{v}{\sqrt{2 \Lambda^2 + v^2 + w^2}}\rho^0_3\crn
		&\simeq& s_\al \varphi^0_2+c_\al \sigma^0_{23}.\eea It is noteworthy that the doublet $(G^+_X,G^0_Y)$---which is composed of two scalar doublets $(\varphi^+_1,\varphi^0_2)$ and $(\sigma^+_{13},\sigma^0_{23})$--- is the Goldstone bosons associated with the vector gauge boson doublet $(X^+,Y^0)$. The Goldstone boson equivalence theorem states that at high energy the experiments which are designed for searching a doublet scalar dark matter, e.g. doublet inert Higgs in scotogenic setup or doublet slepton in supersymmetry, indeed reveal a signature of the doublet vector dark matter. This is because at high energy collider the longitudinal component of vector dark matter is dominantly produced, which is identical to the relevant Goldstone boson mode which we look for.

\subsection{Dark matter observables}

The dark matter candidate must be colorless and electrically neutral. Additionally, it must be the lightest among the odd fields, which is stabilized by matter parity conservation. Depending on the parameter space, the dark matter candidate of the model includes either a singlet pseudo-Dirac fermion $N_{1}$ or $N_2$, or a doublet pseudo-Dirac fermion $E_1$ or $E_2$, or a doublet gauge boson $A_6$ or $A_7$, or a singlet scalar $H'_1$ or $A'_1$, or a doublet scalar $H'_2$ or $A'_2$. It is noted that the normal 3-3-1-1 model contains only singlet fermion and singlet scalar candidates \cite{d1,d1e1}. Furthermore, the doublet vector candidate in the normal 3-3-1-1 model is not separated in mass, hence possessing a large scattering cross-section with nuclei via $Z$ exchange, already ruled out by the direct detection \cite{ddz}. The doublet scalar and doublet fermion candidates are really predicted by this universal 3-3-1-1 model. Although they are well studied in the literature, such candidates at TeV are newly governed by the new gauge and Higgs bosons of the universal 3-3-1-1 model.\footnote{Doublet scalar candidate may be that in the scotogenic mechanism, the inert Higgs doublet model, even sneutrino in supersymmetry. Doublet fermion candidate may be that in the vectorlike lepton doublet model, the model with mirror fermions, and even Higgsino in supersymmetry.} 

We are interested in the doublet vector candidate $A_{6,7}$, uniquely predicted by this universal 3-3-1-1 model. Indeed, such candidate was studied in \cite{srz,dltt} but together with a $Z_2$ dark parity input by hand. Besides the issue of the original nature of the candidate and $Z_2$, the unitarity of the theory is broken at TeV scale. It was well established that the gauge completion theories, such as the gauge-Higgs unification \cite{fairlie,hosotani}, the 3-3-1 model \cite{331m1,331m2,331m3,331r0,331r1,331r2,331r3}, and even gauge extensions containing $SU(3)_L$ as subgroup \cite{dvali}, actually contain such a vector doublet, solving the unitarity question. But, the stability and/or mass splitting of vector dark matter in these approaches remain unsolved. A matter parity may be implemented in the normal 3-3-1-1 model but the vector candidate still degenerate in mass encounters a large direct detection cross-section, as mentioned. 

The vector candidate mass splitting can be done in the trinification model above due to the mixing between the left ($A_{6L}$ and $A_{7L}$) and right ($A_{6R}$ and $A_{7R}$) sectors as well as the contribution of scalar bi-sextet VEVs. By the bi-sextet VEV alignment, $A_{7R}$ is the lightest field responsible for dark matter in \cite{trinifi9}, which belongs to a singlet vector dark matter. Alternatively, the right sector $A_{6R}$ and $A_{7R}$ is integrated out in the universal 3-3-1-1 model, as ascertained. The $A_{6}\equiv A_{6L}$ and $A_7\equiv A_{7L}$ mass splitting is solved in the universal 3-3-1-1 model due to the presence of a scalar sextet, which necessarily determines the matter parity and the right-handed neutrino mass scale. The lightest field $A_7$ predicted is just doublet vector dark matter. Let us call the reader's attention to a scotogenic gauge mechanism in which neutrino mass is radiatively generated by a doublet vector candidate whose mass is separated by interacting with two Higgs doublets confined in an octet of $SU(3)_L$ \cite{dongthao}. 

\begin{figure}[h]
\bc
\includegraphics[scale=0.8]{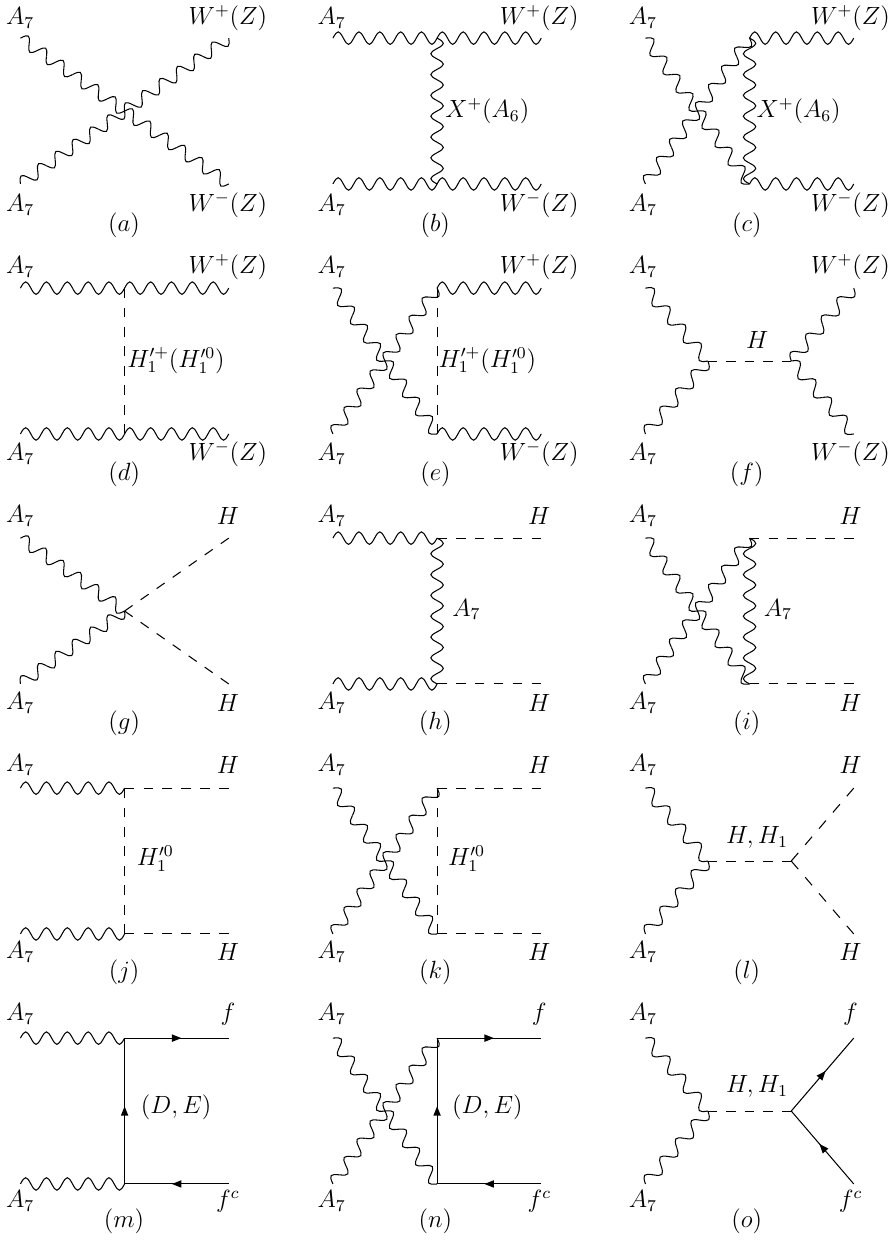}
\caption[]{\label{fig2} Annihilation processes of the doublet vector dark matter $A_7$ to the standard model particles which set the relic abundance.}
\ec
\end{figure}
In the early universe, the dark matter $A_7$'s annihilate to the standard model particles through channels, such as $A_7A_7\to W^+W^-, ZZ, HH, ff^c$, where $f$ denotes usual fermions, which set the dark matter relic density. Such processes are described by Feynman diagrams as depicted in Fig. \ref{fig2}. Notice that each pair of diagrams, say $(b,c)$, $(d,e)$, $(h,i)$, $(j,k)$, and $(m,n)$, are $(t,u)$-channel contributions of the same process. As obtained, the fields $H'^\pm_1\simeq \eta^\pm_3$ and $H'^0_1=(H'_1+i A'_1)\sqrt{2}\simeq \rho^0_3$ are the charged and neutral singlet odd scalars, while the fields $H\simeq c_\beta R_1+s_\beta R_2$ and $H_1\simeq -s_\beta R_1+c_\beta R_2$ for $t_\beta=v/u$ are the usual and new Higgs bosons, respectively. Particularly, the new Higgs fields $H_{2,3,4}$ that are related to $R_{3,4,6}$ may also contribute to the diagram $(l)$, but they are assumed to be radically heavier than the new Higgs field $H_1$, for which their effects are suppressed. The fields $(D,E)$ are new quarks and leptons that appropriately couple to usual fermions ($f$) via $A_7$. Although various processes potentially contribute to the dark matter relic density, the unitarity bounds demand that the corresponding cross sections are proportional to $1/m^2_{A_7}$ as the normal WIMP does. However, because the dark matter mass is at or above TeV, such cross sections are generically small, which translate to an overpopulating density. Fortunately, the dark matter annihilation is substantially enhanced by $s$-channel $H_1$-exchange processes at the $H_1$ mass resonance via diagrams ($l$) and ($o$), which determine the dark matter relic density. Let us see.  

{\it Unitarity constraint.} The processes $A_7A_7\to HH,ff^c$ must yield a cross section proportional to $1/m^2_{A_7}$, which need not necessarily be considered. However, the remaining processes $A_7A_7\to W^+W^-,ZZ$ may cause troublesome, since individual contributions to each cross section are enhanced with the energy scaling, which potentially break the unitarity bound at high energy. Let us consider the process $A_7A_7\to W^+W^-$. The process $A_7A_7\to ZZ$ can be similarly computed, reaching the same conclusion. The process of interest comes from contributions of six diagrams ($a,b,c,d,e,f$) in Fig. \ref{fig2}. Label Lorentz indices and momenta of incoming and outgoing particles such as $A_{7\mu}(p_1)+A_{7\nu}(p_2)\to W^+_\al (k_1)+W^-_\beta (k_2)$. To investigate enhancement of amplitudes with energy scale, it is conveniently worked in the center-of-mass frame associated with two incoming dark matter, i.e. $p_1=(E,\vec{p})$ and $p_2=(E,-\vec{p})$, thus $k_1=(E,\vec{k})$ and $k_2=(E,-\vec{k})$ due to the conservation of energy and momentum. Since the $A_7$ dark matter is nonrelativistic $v\sim 10^{-3}c$, we have $E\simeq m_{A_7}$ and $\vec{p}\simeq m_{A_7}\vec{v}$, thus collision energy $\sqrt{s}=2E\simeq 2m_{A_7}$ and $|\vec{k}|=\sqrt{E^2-m^2_W}\simeq m_{A_7}$. Because of $m_{A_7}\gg m_W$, the products $W$'s are largely boosted in dark matte annihilation, determined through their longitudinal polarization components,
\be \ep_\al (k_1) = \fr{k_{1\al}}{m_W}+\mathcal{O}\left(\fr{m_W}{m_{A_7}}\right),\hs \ep_\beta (k_2) = \fr{k_{2\beta}}{m_W}+\mathcal{O}\left(\fr{m_W}{m_{A_7}}\right),\ee which mainly contribute to the amplitudes. Let us stress that dark matter polarizations $\ep_\mu(p_1), \ep_\nu(p_2)$ are proportional to unity, obeying Lorentz conditions, $p^\mu_1\ep_\mu(p_1)=0=p^\nu_2\ep_\nu(p_2)$. Additionally, $p_1+p_2=k_1+k_2$, $p_1k_1=p_2k_2$, and $m_{W,Z,H}\ll m_{A_6,A_7}\sim m_X\sim m_{H'^-_1,H'^0_1}$ (still ensuring the dark matter to be lightest).          
The interaction Lagrangian that governs the dark matter annihilation process $A_7A_7\to W^+W^-$ is obtained by
\bea \mathcal{L}_{\mathrm{int}} &\supset& \fr{g^2}{4}(g^{\mu\nu}g^{\al\beta}+g^{\mu\beta}g^{\al\nu}-2g^{\mu\al}g^{\nu\beta})A_{7\mu}A_{7\nu}W^+_\al W^-_\beta \crn
&&+\left(-\fr{ig}{2} L^{\mu\nu\al} W^+_\mu A_{7\nu} X^-_\al -\fr{ig^2u}{4} W^+_\mu A^{\mu}_7 H'^-_1+H.c.\right)\crn
&&+\fr{g^2v^2}{4\sqrt{u^2+v^2}}HA^\mu_7A_{7\mu}+\fr{g^2\sqrt{u^2+v^2}}{2} HW^{+\mu}W^-_\mu, \eea where $L^{\mu\nu\al}=g^{\mu\nu}(p_1-p_2)^\al+g^{\nu\al}(p_2-p_3)^\mu+g^{\al\mu}(p_3-p_1)^\nu$ depends on the momenta of $W^+_{\mu}(p_1),A_{7\nu}(p_2), X^-_{\al}(p_3)$ that presumably go into the vertex.   
The amplitudes producing longitudinal $W^\pm$ components according to the relevant diagrams are computed as
\bea && i M(a)=\fr{ig^2}{4m^2_W}\ep_\mu (p_1)\ep_\nu (p_2)\left(2g^{\mu\nu}k_1 k_2 - k_1^\mu k_2^\nu -k_1^\nu k_2^\mu\right)+\mathcal{O}\left(\fr{m^2_W}{m^2_{A_7}}\right),\\
&& i M(b) = \fr{ig^2}{4m^2_W}\ep_\mu (p_1)\ep_\nu (p_2)\left[k^\mu_1 k^\nu_2+g^{\mu\nu}(k^2_2-2p_2k_2)\right]+\mathcal{O}\left(\fr{m^2_W}{m^2_{A_7}}\right),\\
&& i M(c) = \fr{ig^2}{4m^2_W}\ep_\mu (p_1)\ep_\nu (p_2)\left[k^\nu_1 k^\mu_2+g^{\mu\nu}(k^2_2-2p_1k_2)\right]+\mathcal{O}\left(\fr{m^2_W}{m^2_{A_7}}\right),\\
&& i M(d)=\fr{-i g^4u^2}{16 m^2_W}\ep_\mu (p_1)\ep_\nu (p_2)\fr{k^\mu_1 k^\nu_2}{m^2_{A_7}+m^2_{H'^-_1}}+\mathcal{O}\left(\fr{m^4_W}{m^4_{A_7}}\right)\sim 1,\\
&& i M(e)=\fr{-i g^4u^2}{16 m^2_W}\ep_\mu (p_1)\ep_\nu (p_2)\fr{k^\mu_2 k^\nu_1}{m^2_{A_7}+m^2_{H'^-_1}}+\mathcal{O}\left(\fr{m^4_W}{m^4_{A_7}}\right)\sim 1,\\
&& i M(f)=\fr{-i g^2}{4}\ep_\mu (p_1)\ep_\nu (p_2)\fr{k^\mu_1 k^\nu_2}{m^2_{A_7}}+\mathcal{O}\left(\fr{m^4_W}{m^4_{A_7}}\right)\sim 1. 
 \eea The amplitudes $(d,e,f)$ are proportional to 1 due to $k_{1,2}\sim m_{A_7}$, satisfying the unitarity. By contrast, each of amplitudes $(a,b,c)$ is proportional to $m^2_{A_7}/m^2_W$, i.e. $s/m^2_W$, at high energy. Hence, $M(a)$, $M(b)$, and $M(c)$ separately violate the unitarity bound (i.e., the amplitude must be smaller than a constant). However, summing over them,
 \be M(a)+M(b)+M(c)\sim 1,\ee the leading terms add up to zero. Hence, the unitarity is obeyed by the process. In summary, the relevant annihilation cross section $\langle \sigma v_{\mathrm{rel}}\rangle_{A_7A_7\to W^+W^-} \simeq |\sum_{i=a,b,c,d,e,f} M(i)|^2/(32\pi m^2_{A_7})$ is proportional to $1/m^2_{A_7}$, as expected.  

{\it Perturbative regime.} The 3-3-1-1 model with arbitrary embedding of electric charge possesses an electric charge operator of the form $Q=T_{3L}+ b T_{8L}+X$.\footnote{$b$ is an arbitrary embedding coefficient, sometimes labelled $\beta$ in the literature.} This translates to a gauge coupling matching \cite{donglong}
\be \fr{1}{e^2}=\fr{1}{g^2}+\fr{b^2}{g^2}+\fr{1}{g^2_X}. \ee With the aid of $e=g s_W$, we obtain 
\be s^2_W=\fr{g^2_X}{g^2+(1+b^2)g^2_X}<\fr{1}{1+b^2}.\ee This gives rise to a Landau pole, $M$, at which $s^2_W(M)=\fr{1}{1+b^2}$, or $g_X(M)=\infty$. Hence, the present model is valid only if $M$ is larger than the trinification (or dark grand unification) scale, i.e. $M>\Delta$. We have $|b|<\cot_W \simeq 1.825$, where the last value takes $s^2_W\simeq 0.231$ at the weak scale. That said, the model has a Landau pole at the weak scale $M\sim 246$ GeV, if $|b|\to 1.825$. Even if taking $|b|=\sqrt{3}$, or $s^2_W<1/4$, the corresponding model has a low Landau pole, $M\sim 4$--5 TeV \cite{landaupole}. This corresponds to the case of the minimal 3-3-1 model and its variants. Our proposal predicts $b=1/\sqrt{3}$, or $s^2_W<3/4$, which is two times the value predicted by a typical GUT, i.e. $s^2_W=3/8$, at GUT scale. Indeed, the current model predicts a Landau pole beyond the Planck scale. This proves that the dark trinification is viable, reducing to a universal 3-3-1-1 model responsible for dark matter, being testable at TeV scale. Additionally, as running from the trinification scale by RGE's all the gauge interactions in the universal 3-3-1-1 model are perturbative, close to those in the standard model. Furthermore, the scalar and Yukawa couplings must be perturbative too. Hence, the dark matter annihilation governed by this model is typically
\be \langle \sigma v_{\mathrm{rel}}\rangle \simeq \fr{\sum_f |M_{fi}|^2}{32\pi m^2_{A_7}}\sim \fr{g^4}{32\pi m^2_{A_7}}\sim 1\ \mathrm{pb}\left(\fr{800\ \mathrm{GeV}}{m_{A_7}}\right)^2,\ee where $i$ is a pair of initial dark matter, while $f$ is SM products. Above, we take the magnitude of dominant amplitudes as $|M_{fi}|\sim g^2$ similar to the standard model, which can be verified by directly evaluating the amplitudes of processes from Fig. \ref{fig2} with the aid of Feynman rules. Hence, the dark matter candidate with mass at or above TeV negligibly contributes to the annihilation cross section $\langle \sigma v_{\mathrm{rel}}\rangle \simeq 1$~pb as desirable, unless.               

\begin{figure}[h]
\bc
\includegraphics[scale=1]{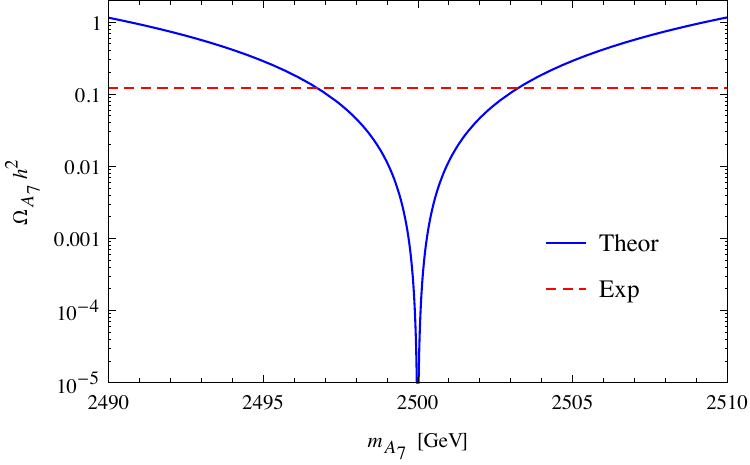}
\caption[]{\label{fig3} Doublet vector dark matter abundance plotted as a function of its mass, where the horizontal dashed line is measured by experiment \cite{pdg}.}
\ec
\end{figure}

{\it Relic density governed by $H_1$ resonance.} Only $s$-channel diagrams lead to a mass resonance that sets the correct relic density. Looking at Fig. 2, the dark matter annihilation is strongly enhanced by $(l,o)$ due to exchanges of $H_1$. Note that the dark matter mass is at TeV, the contribution of $s$-channel diagrams by $H$ exchanges is suppressed. That said, the total annihilation cross section is dominated by 
\be \langle \sigma v_{\mathrm{rel}}\rangle\simeq \langle \sigma v_{\mathrm{rel}}\rangle_{A_7A_7\to HH}+\langle \sigma v_{\mathrm{rel}}\rangle_{A_7A_7\to tt^c}, \ee via $H_1$ exchanges. It is easily estimated,
\be \langle \sigma v_{\mathrm{rel}}\rangle\simeq \fr{g^4v^4\bar{\la}^2}{384\pi m^2_{A_7}(4m^2_{A_7}-m^2_{H_1})^2}+\fr{g^4t^2_\beta v^4 m^2_t}{48\pi (u^2+v^2)^2(4m^2_{A_7}-m^2_{H_1})^2},\ee where $\bar{\la}=6s_{\beta}c_\beta [\la_1 s^2_\beta - \la_3 c^2_\beta+ \fr 1 2 \la_5(c^2_\beta-s^2_\beta)]$. From Fig. \ref{fig1}, taking $t_\beta=v/u=1$, i.e. $u=v\simeq 174$ GeV, the new physics scale $w\sim\La$ can vary in the range from 1 TeV to 7--8 TeV. Correspondingly, the masses of $A_7$ and $H_1$ that are proportional to $(w,\La)$ are naturally given within this regime. Hence, we assume a typical/benchmark mass for the new Higgs boson $H_1$ as $m_{H_1}=5$ TeV. Further, we take $m_t=173$ GeV, $\bar{\la}=1$, and $g\simeq 0.651$, with the aid of $\al=e^2/4\pi=1/128$ and $e=gs_W$. The dark matter relic density $\Omega_{A_7}h^2\simeq 0.1\ \mathrm{pb}/\langle \sigma v_{\mathrm{rel}}\rangle$ is plotted as a function of the dark matter mass in Fig. \ref{fig3}. It is clear from the figure that the $H_1$ mass resonance, i.e. $m_{A_7}\sim \fr 1 2 m_{H_1}$, is crucial to set the dark matter relic density. Compared with the experimental data, $\Omega_{A_7}h^2\simeq 0.12$ \cite{pdg}, we derive $m_{A_7}=2497$--2503 GeV with the choice of parameters, as given. 
                
It is noteworthy that although the vector dark matter relic density is governed by the new Higgs $H_1$, the dark matter scattering with nuclei in direct detection is dominantly mediated by the usual Higgs $H$, instead. Indeed, the scattering process can be described by the effective interaction, such as
\be \mathcal{L}_{\mathrm{eff}} = -\fr{g^2}{2}\fr{s^2_\beta m_q}{m^2_H} A_7 A_7 \bar{q}q +\fr{g^2}{2}\fr{s^2_\beta m_u}{m^2_{H_1}} A_7 A_7 \bar{u}u -\fr{g^2}{2}\fr{c^2_\beta m_d}{m^2_{H_1}} A_7 A_7 \bar{d}d,\ee which can be obtained from the diagram $(o)$ by integrating $H,H_1$ out. Since, typically, $s_\beta\sim c_\beta$ and $m_{H_1}\gg m_{H}$, the last two terms contributed by $H_1$ are radically smaller than the first term contributed by $H$. Hence, the scattering of $A_7$ with nuclei is governed by the usual Higgs boson, such as 
\be \mathcal{L}_{\mathrm{eff}} \simeq 2\la_q m_{A_7} A_7 A_7 \bar{q}q, \ee where
\be \la_q=-\fr{g^2}{4}\fr{s^2_\beta m_q}{m_{A_7}m^2_H}.\ee This leads to a spin-independent (SI) scattering cross section of the vector dark matter $A_7$ with a nucleon ($p/n$), given by \cite{belanger}
\be \sigma^{\mathrm{SI}}_{A_7-p/n}=\fr{4m^2_{p/n}}{\pi}\la^2_{p/n},\ee in which the $A_7$-nucleon coupling is summed over quark-level contributions times relevant nucleon form factors, such as 
\be \fr{\la_{p/n}}{m_{p/n}}=\sum_{q=u,d,s} f^{p/n}_{Tq}\fr{\la_q}{m_q}+\fr{2}{27}f^{p/n}_{TG}\sum_{q=c,b,t}\fr{\la_q}{m_q}\simeq -0.35\fr{g^2}{4}\fr{s^2_\beta}{m_{A_7}m^2_H},\ee where $f^{p/n}_{TG}=1-\sum_{q=u,d,s}f^{p/n}_{Tq}$, while the values of $f^{p/n}_{Tq}$ are given in \cite{eliss}. With $g=0.651$, $s_\beta=1/\sqrt{2}$, $m_{p/n}=1$ GeV, and $m_H=125$ GeV, we get 
\be \sigma^{\mathrm{SI}}_{A_7-p/n}\simeq \left(\fr{2.5\ \mathrm{TeV}}{m_{A_7}}\right)^2 \times 10^{-46}\ \mathrm{cm}^2, \ee in good agreement with the latest measurement for a dark matter mass at TeV \cite{dmdd}.   

\subsection{Collider signals}

\begin{figure}[h]
\bc
\includegraphics[scale=0.8]{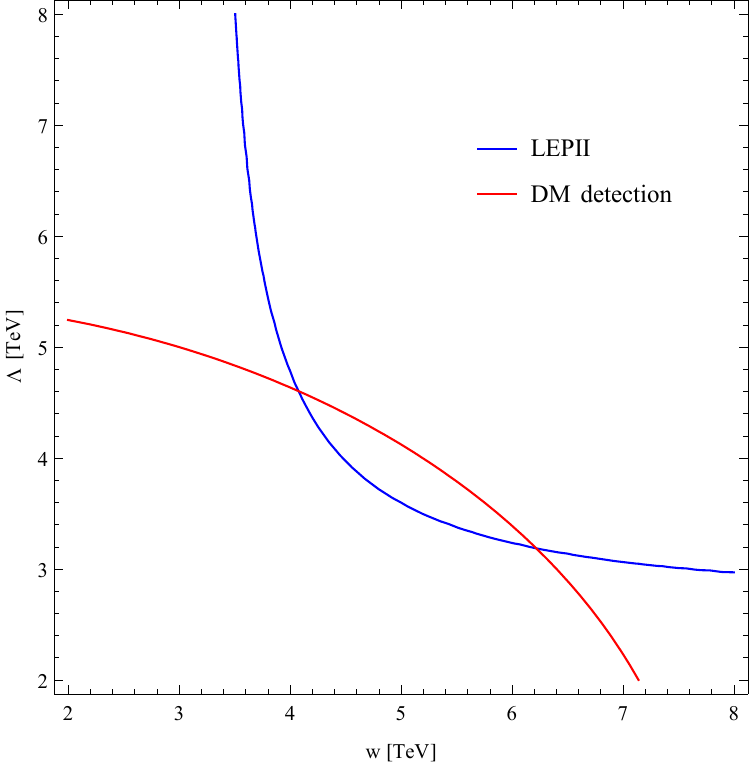}
\caption[]{\label{fig4} LEPII and dark matter direct detection bounds contoured as function of $(w,\La)$.}
\ec
\end{figure}

The LEPII experiment \cite{lep2} studied a signature of new gauge bosons, called $Z_2$ and $Z_3$, which mediate the process such as $e^+e^-\to \mu^+\mu^-$. Since the collider energy of the LEPII is radically smaller than the expected new gauge boson masses, such process can be best described by an effective interaction after integrating these new fields out. First, let us supply the interactions of $Z_{2},Z_{3}$ with leptons,
\be \mathcal{L}\supset \bar{e}\ga^\mu [P_L a^I_L(e) + P_R a^I_R(e)] e Z_{I\mu},\hs I=2,3,\ee where the chiral couplings $a^I_{L,R}$ are family universal, obeying
\bea && a^2_L(e)= \fr{g}{c_W}\left(\fr{c_{2W}}{2\sqrt{3-4s^2_W}}c_\theta -\fr 2 3 c_W t_N s_\theta \right),\\ 
&&a^2_R(e)=\fr{g}{c_W}\left(\fr{c_{2W}}{\sqrt{3-4s^2_W}}c_\theta -\fr 1 3 c_W t_N s_\theta\right),\eea for $Z_2$, while those for $Z_3$ are obtained from $Z_2$ by replacing $(c_\theta,s_\theta) \to (s_\theta, -c_\theta)$. Second, integrating the heavy gauge bosons $Z_2,Z_3$ out, we gain the effective interaction,
\be \mathcal{L}_{\mathrm{LEPII}}=-\left(\fr{[a^2_L(e)]^2}{m^2_{Z_2}}+\fr{[a^3_L(e)]^2}{m^2_{Z_3}}\right)(\bar{e}\ga^\mu P_L e)(\bar{\mu}\ga_\mu P_L \mu)+(LR)+(RL)+(RR),\ee where the last three terms are different from the first one only in chiral structures. Studying such chiral interactions, Ref. \cite{carena} derived a typical bound,
\be \fr{[a^2_L(e)]^2}{m^2_{Z_2}}+\fr{[a^3_L(e)]^2}{m^2_{Z_3}} < \fr{1}{(6\ \mathrm{TeV})^2}. \ee As ascertained, the new gauge boson masses $m_{Z_2}$ and $m_{Z_3}$ as well as the relevant mixing angle $t_{2\theta}$ are all governed by $(w,\La)$ as given in (\ref{mz2z3}) and (\ref{tta}), respectively. Substituting these quantities into the LEPII bound above, we make a contour of it as function of $(w,\La)$ in Fig. \ref{fig4}. We also include a contour for $m^2_{A_7}=(g^2/4)(w^2+2\La^2)=(2.5\ \mathrm{TeV})^2$ as predicted by the dark matter direct detection, for comparison. For this figure, we take $t_N=\sqrt{3}/2$, $g=0.651$, and $s^2_W=0.231$, as given. 
The acceptable regime for $(w,\La)$ should lie above both of the contoured curves, for instance, for $\La=5,4$, and 3 TeV, then $w\geq 4,5$, and 7 TeV, respectively.       

\begin{figure}[h]
\bc
\includegraphics[scale=0.8]{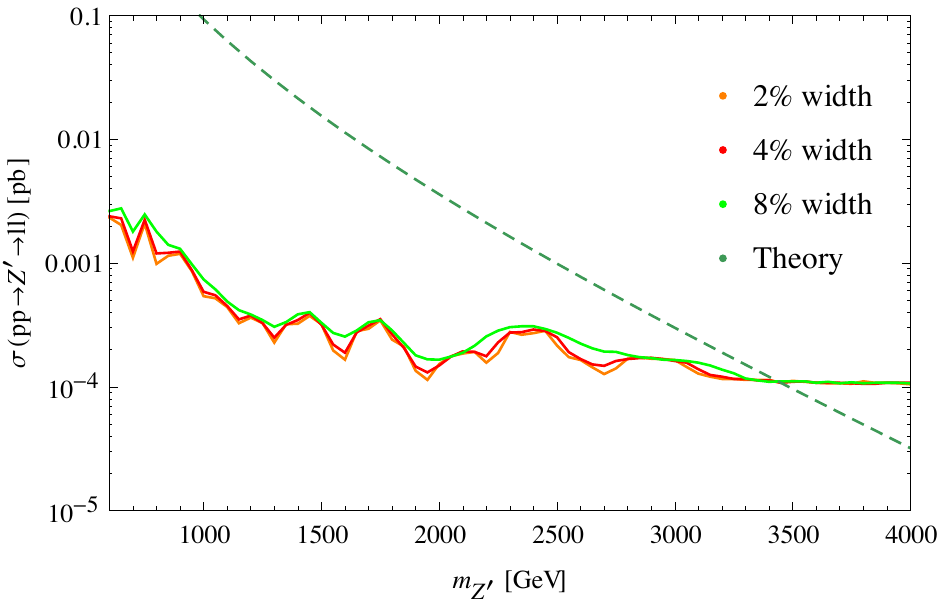}
\caption[]{\label{fig5} Dilepton production cross section plotted as function of the new gauge boson mass at LHC $\sqrt{s}=13$ TeV (dashed curve), where observed limits are deduced at mass resonance for width $\Ga/m=2\%$, $4\%$, and $8\%$ according to ATLAS result \cite{atlas}.}
\ec
\end{figure}

At the LHC, the new gauge bosons $Z_{2,3}$ can be produced on shell and then decay to a dilepton ($ee^c$) or dijet ($qq^c$) signal. The corresponding cross section can be computed by narrow width approximation, such as
\be \sigma(pp\to Z'\to ff^c)=\fr 1 3 \sum_q \fr{dL_{qq^c}}{dm^2_{Z'}}\hat{\sigma}(qq^c\to Z')\mathrm{Br}(Z'\to ff^c),\ee where $Z'\equiv (Z_2,Z_3)$, $q=(u,d)$, and $f=(e,u,d)$. The luminosity $dL_{qq^c}/dm^2_{Z'}$ is given in~\cite{lhcl}. The partonic peak cross section and branching decay ratio $\mathrm{Br}(Z'\to ff^c)=\Ga(Z'\to ff^c)/\sum_{f'}\Ga(Z'\to f'f'^c)$ are obtained by 
\bea && \hat{\sigma}(qq^c\to Z')=\fr{\pi g^2}{12c^2_W}[(g^{Z'}_V(q))^2+(g^{Z'}_A(q))^2],\\
&&\Ga(Z'\to f'f'^c)=\fr{g^2 m_{Z'}}{48\pi c^2_W}[(g^{Z'}_V(f'))^2+(g^{Z'}_A(f'))^2],\eea where $f'=(u,d,e,\nu)$ is all standard model fermions including neutrinos. The couplings of $Z'=(Z_2,Z_3)$ with quarks are computed by
\bea && g_V^2(u)=\fr{3-8s^2_W}{6\sqrt{3-4s^2_W}}c_\theta -\fr{c_W t_N s_\theta}{3},\hs g_A^2(u)=\fr{1}{2\sqrt{3-4s^2_W}}c_\theta +\fr{c_W t_N s_\theta}{3}\\
&& g_V^2(d)=\fr{3-2s^2_W}{6\sqrt{3-4s^2_W}}c_\theta -\fr{c_W t_N s_\theta}{3},\hs g_A^2(d)=\fr{1-2s^2_W}{2\sqrt{3-4s^2_W}}c_\theta + \fr{c_W t_N s_\theta}{3},\eea for $Z_2$, while for $Z_3$ one replaces $(c_\theta,s_\theta)\to (s_\theta,-c_\theta)$. The couplings of $Z_{2,3}$ with leptons are given by $g^I_{V/A} = (-c_W/g)(a^I_L\pm a^I_R)$, respectively, where $a^I_{L,R}$ of charged leptons are given above, while those of neutrinos are $a^I_L(\nu)=a^I_L(e)$ and $a^I_R(\nu)=0$. Generically, the LHC bound would apply for the lighter particle between $Z_{2,3}$. Let us consider a $w,\La$ relation so that $Z_3$ and $Z_2$ (exactly $Z'$ and $C$ as in the gauge boson mass sector) are decoupled, i.e. $t_{2\theta}\sim (2\La^2-w^2)/(w,\La)^2\to 0$, thus $2\La^2\simeq w^2$. In this case, $Z_2$ which is similar to that of the corresponding 3-3-1 model is relevant to the LHC search. We plot the dilepton production cross section as in Fig. \ref{fig5} at the LHC $\sqrt{s}=13$ TeV, corresponding to an integrated luminosity of 139 $\mathrm{fb}^{-1}$ \cite{atlas}. The LHC reveals a negative result, correspondingly making a bound for $Z'$ mass as $m_{Z'}>3.4$ TeV, appropriate to the LEPII and dark matter detection bounds above. 
It is noted that the dijet search is less sensitive than that of lepton. Hence, the relevant bound is weaker than the given dilepton bound, which will not further be considered.  

Finally, the LHC can produce a dark matter signal given in terms of missing energy. The vector dark matter behaves as the Goldstone boson mode at high energy, as mentioned. Therefore, the former searches for such a doublet scalar dark matter can be applied to our case, without change. The LHC studied a pair production of dark matter slepton before decaying to a dilepton signal plus missing energy, making a typical bound as $m_{A_7}>700$ GeV \cite{dvsearch,dvsearch1}, which is obviously satisfied due to the LEPII and direct detection bounds.     

\section{\label{con} Conclusion}

We have shown that $E_6$ and trinification can act as a dark grand unification, which unifies dark matter and normal matter in a nontrivial manner in a grand unified theory. Their breaking naturally leads to dark matter stability and neutrino mass generation at low energy. The $E_6$ unification broken implies a trinification-type theory at TeV scale, containing a vast of dark fields. However, the trinification broken manifestly reveals a universal 3-3-1-1 model at TeV scale, which is recognized for the first time and very predictive. First of all, the issue of tree-level FCNCs in the normal 3-3-1 and 3-3-1-1 model is solved. The neutrino mass is produced by a seesaw. The matter parity is automatically conserved at low energy. Dark matter candidates include a singlet and doublet scalar, a singlet and doublet fermion, and a doublet vector. All of them may be stabilized, responsible for dark matter. The direct detection is easily evaded, since doublet candidates are separated in mass. Considering the doublet vector as dark matter, we derive that its mass is 2.5 TeV, appropriate to precision electroweak measurement, dark matter direct detection, LEPII, and LHC bounds.       

In the scotogenic setup, the mass splitting of doublet neutral inert scalar component is crucial to set radiative neutrino mass generation. Given that the scotogenic scheme is of a gauge completion such that the inert scalar doublet becomes the Goldstone boson of a gauge vector doublet. The question is that may the radiative neutrino mass generation be mediated by the doublet vector instead of that before the gauge completion? The 3-3-1-1 model might reveal such a scotogenic gauge mechanism as worth exploring, giving a contribution comparable to the tree-level seesaw contribution. However, the theory also gives rise to a coupling (identical to charged lepton coupling) of the active neutrino with a dark doublet neutral lepton and a dark singlet scalar inducing scotogenic neutrino masses appropriate to experiment, given that the second and third family right-handed neutrinos are light, namely they obtain Majorana masses at the weak (GeV) scale and GeV (10 MeV) scale for $u\sim v$ ($u/v\sim 10$), respectively. In all cases, the Majorana mass of the first family right-handed neutrino is above TeV.

\section*{Ackowledgements}    

This research is funded by Vietnam National Foundation for Science and Technology Development (NAFOSTED) under grant number 103.01-2023.50.

\end{document}